\documentclass[journal]{IEEEtran}
\ifCLASSINFOpdf
\else
\fi

\usepackage{amsmath}%
\usepackage{amsfonts}%
\usepackage{amssymb}%
\usepackage{graphicx}
\usepackage{url}
\usepackage{booktabs}
\usepackage{multirow}
\usepackage{placeins}
\usepackage{comment}
\usepackage{todonotes}
\usepackage{url}
\usepackage{pifont}
\usepackage{subcaption}

\usepackage{bm}
\usepackage{textcomp}

\usepackage{math/math_defs}

\usepackage{xcolor}
\usepackage{enumitem}

\usepackage{array}
\newcolumntype{P}[1]{>{\centering\arraybackslash}p{#1}}

\def\vsseca{0mm}

\usepackage{caption}
\DeclareMathAlphabet{\pazocal}{OMS}{zplm}{m}{n}
\newcommand{\unif}{\pazocal{U}}
\usepackage[thinc]{esdiff}
\begin{document}
%
\title{Study of Pre-processing Defenses against Adversarial Attacks on State-of-the-art Speaker Recognition Systems}
%
%
%

\author{Sonal Joshi,~\IEEEmembership{Student Member,}
        Jesús Villalba,~\IEEEmembership{Member,~IEEE,}
        Piotr Żelasko,~\IEEEmembership{Member,~IEEE,}
        Laureano Moro-Velázquez,~\IEEEmembership{Member,~IEEE,}
        and~Najim Dehak,~\IEEEmembership{Senior Member,~IEEE}
        
\thanks{All authors are associated with the Department
of Electrical and Computer Engineering, and the Center for Language and Speech Processing (CLSP), Johns Hopkins University, Baltimore,
MD, 21218 USA  }
\thanks{Jesús Villalba, Piotr Żelasko and Najim Dehak are also affiliated to the Human Language Technology Center of Excellence, Johns Hopkins University, Baltimore, MD, 21218, USA}
\thanks{This project was supported by DARPA Award HR001119S0026-GARD-FP-052}
\thanks{Manuscript received January xx, 2021}}

%
%

\markboth{Journal of \LaTeX\ Class Files,~Vol.~XX, No.~XX, January~2021}%
{Shell \MakeLowercase{\textit{et al.}}: Bare Demo of IEEEtran.cls for IEEE Journals}
%



\maketitle


\begin{abstract}
Adversarial examples to speaker recognition (SR) systems are generated by adding a carefully crafted noise to the speech signal to make the system fail while being imperceptible to humans. Such attacks pose severe security risks, making it vital to deep-dive and understand how much the state-of-the-art SR systems are vulnerable to these attacks. Moreover, it is of greater importance to propose defenses that can protect the systems against these attacks. Addressing these concerns, this paper at first investigates how state-of-the-art x-vector based SR systems are affected by white-box adversarial attacks, i.e., when the adversary has full knowledge of the system. x-Vector based SR systems are evaluated against white-box adversarial attacks common in the literature like fast gradient sign method (FGSM), basic iterative method (BIM)--a.k.a. iterative-FGSM--, projected gradient descent (PGD), and Carlini-Wagner (CW) attack. To mitigate against these attacks, the paper proposes four pre-processing defenses. It evaluates them against powerful adaptive white-box adversarial attacks, i.e., when the adversary has full knowledge of the system, including the defense. The four pre-processing defenses--viz. randomized smoothing, DefenseGAN, variational autoencoder (VAE), and Parallel WaveGAN vocoder (PWG) are compared against the baseline defense of adversarial training. Conclusions indicate that SR systems were extremely vulnerable under BIM, PGD, and CW attacks. Among the proposed pre-processing defenses, PWG combined with randomized smoothing offers the most protection against the attacks, with accuracy averaging  93\%  compared  to  52\%  in  the  undefended system and an absolute improvement $>$ 90\% for BIM attacks with  $L_\infty>0.001$ and CW attack.

\end{abstract}


%
\IEEEpeerreviewmaketitle

\vspace{\vsseca}
\section{Introduction}
%
%
%
%

\begin{figure*}
    \centering
    \includegraphics[width=0.7\textwidth]{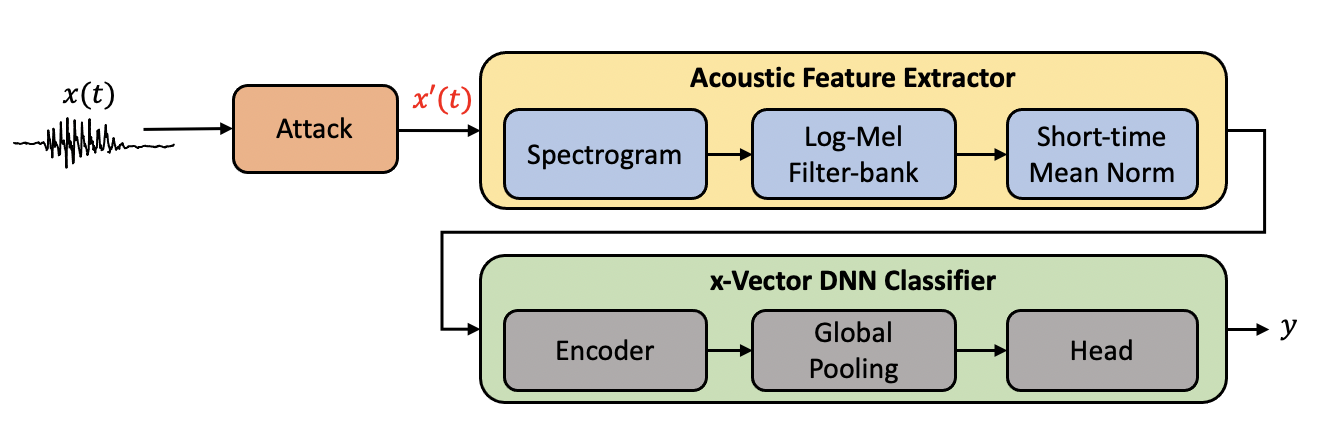}
    \caption{x-Vector Speaker Classification Pipeline. Here, $\xvec^\prime$ is adversarial sample of benign waveform $\xvec$ (with without attack gives classifier output as speaker label $y^{\mathrm{benign}}$) such that the classifier output is $y^{\mathrm{adv}}$ (or in short $y$) such that $y \neq y^{\mathrm{\mathrm{benign}}}$}
    \label{fig:pipeline}
    \vspace{-5mm}
\end{figure*}

\IEEEPARstart{R}{esearch} in speaker recognition (SR) has been undertaken for several decades, reaching great performance. However, recently, research has shown that these systems are subject to threats. We can broadly classify threats against SR into spoofing and adversarial attacks~\cite{das_attackers_2020}. Spoofing attacks--i.e., attacks using voice replay, conversion, and synthesis--have been extensively studied due to ASVspoof challenges~\cite{wu2015spoofing}~\cite{Todisco2019}, and several countermeasures have been proposed~\cite{das_attackers_2020}~\cite{kamble2020advances}~\cite{Lai2019}. However, quite recently, a new attack genre, termed adversarial--introduced by the computer vision (CV) community--has shown to be fatal for SR~\cite{villalba2020x}. 
Adversarial attacks on neural networks are carried out by malicious inputs called adversarial examples. Such adversarial examples are created by slightly modifying the input to a neural network by adding a carefully crafted noise. This noise is generally not perceptible to humans but makes the neural network fail and give incorrect predictions~\cite{szegedy-iclr14}. As the pioneering work on adversarial attacks was in the area of image classification~\cite{szegedy-iclr14}, attacks to computer vision systems have been extensively studied~\cite{Goodfellow2015}~\cite{carlini-16}~\cite{Kurakin2017}~\cite{Dong2019}. These attacks have been, later, extended to other modalities like video~\cite{hao2020adversarial} 
and speech~\cite{vaidya2015cocaine}~\cite{iter-17}~\cite{Cisse2017}~\cite{carlini-18}~\cite{kreuk-icassp18}~\cite{gong-18}. For the latter, more research focus has been on automatic speech recognition (ASR) systems ~\cite{wang2020adversarial} which convert speech to text, and comparatively, the SR systems~\cite{abdullahsok2020} have been lesser studied. With growing applications of SR in forensics, smart home assistants, authentication, etc., it is extremely vital to study the threats against them and propose remedial measures.

This paper focuses on adversarial attacks against speaker recognition systems and proposes four pre-processing defenses with the following major contributions: 
\begin{itemize}
    \item We show extensive analysis of adversarial attacks at various strengths on the state-of-the-art speaker recognition system (x-vector) and show its vulnerability to adversarial attacks. 
    \item We propose four pre-processing defenses, viz. randomized smoothing (Section \ref{sec:smooth}), Defense-GAN (Section \ref{sec:defgan}), variational auto-encoder (VAE) (Section \ref{sec:vae}), and Parallel-WaveGAN (PWG) vocoder (Section \ref{sec:wavegan}). While the first two pre-processing defenses are inspired by the computer vision literature, the last two (VAE and PWG) are speech-specific defenses newly proposed in this paper. 
    \item We compare the pre-processing  defenses against the baseline of FGSM and PGD adversarial training, which has already been proposed in the literature. 
    \item We evaluate the robustness of the defenses against strong end-to-end adaptive white-box attacks, where both, the speaker recognition model and defense are made available to the adversary. To the best of our knowledge, it is the first study that performs such extensive analysis with attack strengths ranging from 0.0001 to 0.2. We also evaluate the robustness of the defenses against approximated adaptive white-box and transfer black-box attacks.
    \item We show that while the x-vector system has inherent robustness against some attacks like transfer universal attacks and low $L_\infty$ FGSM attacks, defenses like the PWG vocoder and VAE help the x-vector to withstand stronger attacks.
\end{itemize}

The rest of the paper is organized as follows: Section \ref{sec:prior} gives an overview of existing literature, Section \ref{sec:xvec}
describes the x-vector SR network; Section \ref{sec:threat} describes the considered threat model; Section \ref{sec:adv_attacks} formally introduces adversarial attack algorithms; Section \ref{sec:defenses} describes defense methods against the adversarial attacks; Section \ref{sec:exp} shows the experimental setup, while the results are discussed in Section \ref{sec:results}; and finally, Section \ref{sec:conclusions} gives concluding remarks.

\section{Prior work}
\label{sec:prior}

Adversarial attacks in speech systems were first studied on ASR~\cite{vaidya2015cocaine}. In contrast, there are fewer investigations for adversarial attacks on SR systems. Hence, although our work focuses on adversarial pre-processing defenses against SR, we review defenses as well as attacks for both ASR and SR as they fall under the same broad area of speech.

\subsection{Adversarial attacks on speech systems}

Vaidya et al.~\cite{vaidya2015cocaine} were the first to show the existence of  hidden voice commands that are recognized by ASR but are considered as simply a noise by humans. Building on this work, Carlini et al.~\cite{carlini2016hidden} showed the existence of stronger attacks, which are imperceptible to humans. Cisse et al.~\cite{Cisse2017} proposed the Houdini attack, an attack that can fool any gradient-based learning neural network by generating an attack tailored for the final performance measure of the task considered allowing it to attack a wide range of prediction models rather than just classification models commonly found in the literature. This attack approximately doubles the Word Error Rate (WER) of the DeepSpeech~\cite{amodei2016deep} end-to-end model. Iter et al.~\cite{iter-17} attacked WaveNet~\cite{oord2016wavenet} using the fast gradient sign method~\cite{goodfellow_generative_2014} and the fooling gradient sign method~\cite{szegedy-iclr14}.  Carlini and Wagner~\cite{carlini-18} proposed targeted white-box iterative optimization-based adversarial attack (henceforth referred to as Carlini-Wagner attack), making DeepSpeech transcribe any chosen phrase with a 100\% success rate.
Yuan et al.~\cite{yuan-18} proposed \textit{Commandersong}--a surreptitious attack to a Kaldi~\cite{povey2011kaldi} based ASR by embedding voice into songs and playing it in the background while being inaudible to the human ear.  Neekhara et al.~\cite{Neekhara2019} proposed a single quasi-imperceptible perturbation called ``universal adversarial perturbation'', which will most likely cause the ASR to fail when added to any arbitrary speech signal. Schonherr et al.~\cite{Schonherr2019} and Qin et al.~\cite{Qin2019} show that psychoacoustic modeling can be leveraged to make the attacks imperceptible.
Some other works~\cite{Qin2019}~\cite{Yakura2019} suggest that physical adversarial attacks, meaning adversarial attacks generated over-the-air using realistic simulated environmental distortions, can also perturb the ASR systems' performance. A more comprehensive overview of adversarial attacks and countermeasures on ASR is presented by Wang et al.~\cite{wang2020adversarial}.

While many studies have been done for ASR, comparatively adversarial attacks on SR have been lesser studied. SR is an umbrella term for two areas: speaker identification or classification (given a voice sample, identify who is speaking), speaker verification (given a voice sample and a claimed identity, decide whether the claim is correct or not). The adversarial attacks research in the SR field started on non-state-of-the-art speaker identification systems. For instance, Kreuk et al.~\cite{kreuk-icassp18} and Gong and Poellabauer~\cite{gong-18} attacked models based on recurrent Long Short Term Memory (LSTM). Recently, a handful of works attacked the state-of-the-art x-vector~\cite{snyder-icassp18} and i-vector~\cite{dehak_ivector} systems. While Li et al.~\cite{Li2020a} attacked Gaussian Mixture Model (GMM) i-vector and x-vector models using FGSM, Xie et al.~\cite{Xie2020} attacked a public pre-trained time-delay neural network (TDNN) based x-vector model by proposing a real-time black-box universal attack. Li et al.~\cite{li2020universal} introduced ``universal adversarial perturbation" for SR systems using generative networks to model the low-dimensional manifold of UAPs. Wang et al.~\cite{wang2020inaudible} proposed an imperceptible white-box psychoacoustics-based method to attack an x-vector model. Villalba et al.~\cite{villalba2020x} conducted a comprehensive study of the effects of adversarial attacks on speaker verification x-vector systems and successfully transfer white-box attacks like FGSM, randomized FGSM, and CW from small white-box to larger black-box systems. Not just SR, even systems with countermeasures for spoofing attacks have shown to be susceptible to adversarial attacks~\cite{zhang2020black}. 
A recent survey study by Abdullah et al.~\cite{abdullahsok2020} provides a good overview of existing research for attacks on both ASR and SR systems. 

\subsection{Countermeasures against adversarial attacks on speech systems}
Given the previous studies, it is of prime importance to protect speech systems against adversarial attacks. According to Wang et al. ~\cite{wang2020adversarial}, countermeasures against adversarial attacks can be classified as
\begin{itemize}
    \item \textit{Proactive defenses}: these defenses entail (re-)training models to make them inherently robust to adversarial samples. An example is adversarial training~\cite{goodfellow2014explaining}~\cite{madry2018towards}, which simply means training the models using adversarial samples generated on-the-fly.
    \item \textit{Reactive defenses}: these defenses do not modify the model but add additional blocks (generally before) to the model. These can be divided into,
    \begin{itemize}
    \item  \textit{Adversarial attack detection}: it acts as a gate before the system under attack. The goal is to reject the adversarial examples before passing them to the system for processing.
    It is a form of denial-of-service.
    \item \textit{Pre-processing}: it reconstructs the input signal with the aim to remove the adversarial perturbation before feeding it to the system~\cite{song2018pixeldefend,samangouei_defense-gan_2018}. 
    \end{itemize}
\end{itemize}

 Adversarial training using fast gradient sign method (FGSM) attacks was first proposed by Goodfellow et al.~\cite{goodfellow2014explaining} and Kurakin et al.~\cite{kurakin2016adversarial} in the CV domain. However, Madry et al.~\cite{madry2018towards} showed that using projected gradient descent (PGD) attacks makes the system more robust. Moving back to the speech domain,
 Wang et al.~\cite{wang2019adversarial} proposed FGSM adversarial training to avoiding over-fitting in speaker verification systems.
 However, the experimentation regarding adversarial robustness is limited.
 Recently, Jati et al.~\cite{jati_adversarial_2020} proposed PGD adversarial training to defend speaker identification. However, they evaluate its robustness only against $L_\infty = 0.002$. Pal et al.~\cite{pal2020adversarial} proposed a defense mechanism based on a hybrid adversarial training using multi-task objectives, feature-scattering, and margin losses. Nevertheless, the caveats of adversarial training are that it generalizes poorly to attack algorithms and threat models unseen during training~\cite{zhang2019limitations}~\cite{laidlaw2020perceptual}, requires careful training hyperparameter tuning, and significantly increases training time. We used adversarial training as our baseline defense. 
 
 Rajaratnam and Kalita~\cite{rajaratnam2018noise} proposed an attack detection method that uses a ``flooding'' score representing the amount of random noise that changes the model's prediction when added to the signal. A threshold over this score decides whether the example is adversarial or not. Although this threshold depends heavily on the training dataset and attack methods, the authors have shown its effectiveness against only one type of attack. This raises concerns about the generalization ability of the method against multiple attack algorithms. Li et al.~\cite{li2020investigating} proposed an attack detection for adversarial attacks on speaker verification using a separate VGG-like binary classifier. However, the method fails for unseen attacks. Yang et al.~\cite{yang2018characterizing} proposed another detection method by exploiting temporal dependency in audio signals. In this method, two ASR outputs--the text recognized when the input is the first \textit{n} seconds of audio and the text corresponding to the first \textit{n} seconds obtained when the input is the entire audio--are compared using a distance measure metric. If the distance is small, it means that the temporal dependency is preserved and the audio is benign. However, such a defense strategy depends on the choice of $n$. Also, it heavily relies on the existence of a robust ASR model, which may not be available for in-the-wild speaker recognition datasets. 
 
 Yuan et al.~\cite{yuan-18} proposed two pre-processing defenses for ASR systems: adding noise and downsampling the input audio. However, they demonstrated the effectiveness of the defenses (in terms of success rate\%) only against \textit{CommanderSong} attack using a test-set that cannot be benchmarked. Yang et al.~\cite{yang2018characterizing} propose three pre-processing defenses, viz. local smoothing using a median filter, down-sampling/ up-sampling, and quantization. However, they showed that defenses are not effective against adaptive attacks. Esmaeilpour et al. ~\cite{esmaeilpour2021} pre-processed the input signal using a class-conditional Generative Adversarial Network (GAN), in the same style as DefenseGAN~\cite{samangouei_defense-gan_2018} in images, for defense against attacks on DeepSpeech and Lingvo based ASRs. Although there are some similarities between our proposed defense and this work, we clearly point out the similarities and differences in Section~\ref{sec:defgan}. Andronic et al. ~\cite{andronic2020mp3} proposed using the lossy MP3 compression algorithm that discards inaudible noise as a pre-processing defense against imperceptible adversarial attacks on speech. However, they study the defense only against FGSM attacks and suggest re-training the ASR system on compressed MP3 data for best results.
Given these issues, our work focuses on studying pre-processing defenses which do not need training using adversarial examples and do not need to know about the attack algorithms used by the adversary. 

\vspace{\vsseca}
\section{x-Vector Speaker Recognition}
\label{sec:xvec}
In this study, we used state-of-the-art \emph{x-vector} style neural architectures~\cite{snyder-is17}~\cite{snyder-icassp18} for the speaker identification task. x-Vector networks are divided into three parts. First, an encoder network extracts frame-level representations from acoustic features such as MFCC or filter-banks. Second, a global temporal pooling layer aggregates the frame-level representation into a single vector per utterance.  Finally, a classification head, a feed-forward classification network, processes this single vector and computes the speaker class posteriors. This network is trained using a large margin cross-entropy loss version such as additive angular margin softmax (AAM-softmax)~\cite{Deng2019}. Different x-vector systems are characterized by different encoder architectures, pooling methods, and training objectives. 
In preliminary experiments, we compared the vulnerability of x-vectors based on ThinResNet34, ResNet34~\cite{zeinali2019but}~\cite{Villalba2019a}, EfficientNet~\cite{tan2019efficientnet} and 
Transformer~\cite{vaswani2017attention} Encoders to adversarial attacks. For the speaker identification experiments, 
we used the output of the x-vector network to make a decision. For the speaker verification experiments, we extracted embedding from the first hidden layer of the classification head. We then calibrated the enrollment and test embeddings into log-likelihood ratios and compared them using cosine scoring.

\vspace{\vsseca}
\section{Adversarial Threat Model}
\label{sec:threat} 
An adversary crafts an adversarial example by adding an imperceptible perturbation to the speech waveform in order to alter the speaker identification decision. Suppose $\xvec\in\real^T$ is the original clean audio waveform of length $T$, also called benign, or bonafide. Let $y^{\mathrm{benign}}$ be its true label and $g(.)$ be the x-vector network predicting speaker class posteriors. An adversarial example of $\xvec$ is given by $\xvec'=\xvec+\deltavec$, where $\deltavec$ is the adversarial perturbation. 

$\deltavec$ can be optimized to produce untargeted or targeted attacks. Untargeted attacks make the system predict any class other than the true class, $g(\xvec')\ne y^{\mathrm{benign}}$, but without forcing to predict any specific target class.  On the contrary, targeted attacks are successful only when the system predicts a given target class $y^{\mathrm{adv}}$ instead of the true class, $g(\xvec')=y^{\mathrm{adv}}\ne y^{\mathrm{benign}}$. In  this  work,  we  consider  untargeted  attacks  since
we observed that it is easier to make them succeed. Thus, we expect that defending from those attacks will be more difficult.

To enforce imperceptibility of the perturbation, some distance metric is minimized or bounded $\mathcal{D}(\xvec,\xvec')<\varepsilon$. Choosing this metric is termed as \emph{threat model} of the attack.
Typically, this is the $L_p$ norm of the perturbation, $\mathcal{D}(\xvec,\xvec')=\left\|\deltavec\right\|_p$
Among the attack algorithms that we considered (described in Section \ref{sec:adv_attacks}), 
FGSM, Iter-FGSM, and PGD attacks limit the maximum $L_\infty$ norm, while Carlini-Wagner minimizes the $L_2$ norm.

We considered three variants of this threat model depending on the knowledge of the adversary:
\begin{itemize}
    \item \textit{White-box attacks}: the adversary has full knowledge of the system, including architecture and parameters. The adversary can compute the gradient of the loss function and back-propagate it all the way to the waveform to compute the adversarial perturbation. In this work, we performed experiments with three attacks in white-box settings: FGSM~\cite{Goodfellow2015} (Section \ref{sec:fgsm}), BIM and PGD ~\cite{Kurakin2017} (Section \ref{sec:bim}), and Carlini-Wagner~\cite{carlini-16} (Section \ref{sec:cw}) attacks. 
    \item\textit{Black-box attacks}: the adversary does not have details about the architecture but can observe the predicted labels of the system given an input. In this work, we consider a particular case of black-box attack, usually known as \emph{transfer-based black-box attack} (Section \ref{sec:universal}). Here, the adversary has access to an alternative white-box model and uses it to create adversarial examples and attack the victim model.
    \item \textit{Grey-box attack}: the adversary has access to some parts of the model but not others. For example, the adversary may know the parameters of the speaker recognition model but not the parameters of the defense.
    \end{itemize}
Depending on the information available to the adversary about the defense, an attack can be categorized as adaptive or non-adaptive~\cite{abdullahsok2020}.
\begin{itemize}
    \item \textit{Non-adaptive attack}: The adversary just uses the speaker recognition system to generate the adversarial sample and does not consider the effect of the defense. We did not consider this kind of attack since they overestimate the efficacy of the defenses. 
    \item \textit{Adaptive attack}: The attack is specifically designed to target a given defense. Tramer et al.~\cite{tramer2020adaptive} strongly encourage authors proposing defenses to focus their robustness evaluation on adaptive attacks as they uncover and target the defense’s weakest links. Following their suggestion, our work solely focuses on adaptive attacks. We considered two types of adaptive attacks:
    \begin{itemize}
        \item \textit{Backward pass differentiable approximation (BPDA)}: this is used when the adversary cannot compute the Jacobian of the defense block $m(x)$, either when the parameters of defense are unknown (black-box defense) or when $m(x)$ is non-differentiable. BPDA~\cite{athalye2018obfuscated} approximates the defense by a differentiable function $q(x)\approx m(x)$. While computing the adversarial sample, we forward pass through the original defense $m(x)$, but we approximate the backward pass using $\nabla_x q(x)$.
        Although slightly inaccurate, BPDA proves useful in constructing an adversarial example.
        \item \textit{End-to-end differentiable (E2ED)}: when both, defense and victim model, are white-box fully differentiable functions, we can exactly compute the gradient of loss function given the input for the full pipeline, including the defense and the speaker recognition system.
     \end{itemize}
     For variational autoencoder and Parallel WaveGAN defenses, we compared both, BPDA and E2ED, adaptive attacks. For DefenseGAN, only BPDA was possible since it is a non-differentiable function.
\end{itemize}

\vspace{\vsseca}
\section{Adversarial attack algorithms}

\label{sec:adv_attacks}
\vspace{-1mm}
\subsection{Fast gradient sign method (FGSM)}
\label{sec:fgsm}
FGSM~\cite{Goodfellow2015} takes the benign audio waveform of length $T$ $\xvec\in\real^T$  and computes an adversarial example $\xvec^\prime$ by taking a single step in the direction that maximizes the misclassification error as 
\begin{align}
    \label{eq:fgsm}
    \xvec^\prime = \xvec + \varepsilon\,\sign(\nabla_\xvec L(g(\xvec),y^{\mathrm{benign}})) \;,
\end{align}
where function $g(\xvec)$ is the x-vector network producing speaker label posteriors, $L$ is categorical cross-entropy loss, $y^{\mathrm{benign}}$ is the true label of the utterance. $\varepsilon$ restricts the  $L_\infty$ norm of the perturbation by imposing $ \lVert \xvec^\prime - \xvec \rVert_{\infty}
\leq \varepsilon$ to keep the attack imperceptible. In other words, the larger the epsilon, the more perceptible are the perturbations and the more effective are the attacks (meaning classification accuracy deteriorates).

\subsection{Basic Iterative Method (BIM) and Projected Gradient Descent (PGD)}
\label{sec:bim}
Basic iterative method (BIM)--a.k.a. iterative FGSM--~\cite{Kurakin2017} takes iterative smaller steps $\alpha$ in the direction of the gradient in contrast to FGSM, which takes a single step
\begin{align}
    \label{eq:iterfgsm}
   \xvec^{\prime}_{i+1} = \xvec + \clip_\varepsilon(
   \xvec^{\prime}_{i} + \alpha\,\sign(\nabla_{\xvec^{\prime}_{i}} L(g(\xvec^{\prime}_{i}), y^{\mathrm{benign}})) -\xvec) \;,
\end{align}
where $\xvec^{\prime}_0=\xvec$ and $i$ is iteration step for optimization. The $\clip$ function assures that the $L_\infty$ norm of perturbation is smaller than $\varepsilon$ after each optimization step $i$. This results in a stronger attack than that of FGSM, however it also takes more time to compute. In our experiments, we used $\alpha=\varepsilon/5$. The number of iterations is chosen heuristically to keep the computational cost under check yet sufficient enough to reach edge of $\varepsilon$ ball~\cite{kurakin2016adversarial}.

Projected Gradient Descent (PGD)~\cite{madry2018towards} is a generalization of BIM for arbitrary $L_p$ norms--usually 1, 2, or $\infty$. Thus, BIM is equivalent to PGD-$L_\infty$.
\begin{align}
    \label{eq:pgd}
   \xvec^{\prime}_{i+1} = \xvec + \mathcal{P}_{p,\varepsilon}(
   \xvec^{\prime}_{i} + \alpha\,\sign(\nabla_{\xvec^{\prime}_{i}} L(g(\xvec^{\prime}_{i}), y^{\mathrm{benign}})) -\xvec) \;,
\end{align}
where $\mathcal{P}_{p,\varepsilon}$ is a projection operator in the $L_p$ 
ball. PGD attack can also consider the option of taking several random initializations for the perturbation $\delta$ and using the one that produces the highest loss.

\vspace{-3mm}
\subsection{Carlini-Wagner attack}
\label{sec:cw}
The Carlini-Wagner (CW) attack \cite{carlini-16} is computed by finding the minimum perturbation $\deltavec$ that fools the classifier while maintaining imperceptibility.
$\deltavec$ is obtained by minimizing the loss,
\begin{align}
    \label{eq:cw}
    C(\deltavec) \triangleq \mathcal{D}(\xvec,\xvec+\deltavec) + c \,f(\xvec+\deltavec)\;,
\end{align}
where, $\mathcal{D}$ is a distance metric, typically $L_2$ norm. By minimizing $\mathcal{D}$, we minimize the perceptibility of the perturbation.

The function $f$ is a criterion defined in such a way that the system fails if and only if $f(\xvec+\deltavec)\le 0$. 
To perform a non-targeted attack in a classification task, we define $f(\xvec^\prime)$ as,
\begin{align}
    f(\xvec^\prime)=\max(g(\xvec^\prime)_{\mathrm{benign}}-\max\{g(\xvec')_j, j\ne \mathrm{benign}\} + \kappa, 0),
\end{align}
where $\xvec^\prime$ is the adversarial sample,  
$g(\xvec')_j$ is the logit prediction for class $j$, and $\kappa$ is a confidence parameter. The attack is successful when $f(\xvec^\prime)\le 0$. This condition is met when, at least, one of the logits of the non-benign classes is larger than the logits of the benign class plus $\kappa$. We can increase the confidence in the attack success by setting $\kappa > 0$.

The weight $c$ balances $\mathcal{D}$ and $f$ objectives. The optimization algorithm consists of a nested loop. In the outer loop,
A binary search is used
to find the optimal $c$ for every utterance. In the inner loop, for each value of $c$, $C(\deltavec)$ is optimized by gradient descend iterations. If an attack fails,  $c$ is increased in order to increase the weight of the $f$ objective over $\mathcal{D}$, and the optimization is repeated. If the attack is successful, $c$ is reduced.

For our experiments, we set the learning rate parameter as 0.001, confidence  $\kappa$ as $0$, the maximum number of inner loop iterations as 10, and the maximum number of halving/doubling of $c$ in the outer loop as 5.

\vspace{-3mm}
\subsection{Transfer universal attacks}
\label{sec:universal}
Universal perturbations are adversarial noises that are optimized to be sample-agnostic~\cite{moosavi2017universal}. Thus, we can add the same universal perturbation to multiple audios and make the classifier fail for all of them.
The algorithm to optimize these universal perturbations is summarized as follows. Given a set of samples $\Xmat=\{\xvec_1,\dots,\xvec_N\}$, we intend to find a perturbation $\deltavec$, such that $\left\|\deltavec\right\|_{p}\le\varepsilon$,
that can fool most samples in $\Xmat$. 
The algorithm iterates over the data in $\Xmat$ gradually updating $\deltavec$. In each iteration $i$, if the current  $\deltavec$ does not fool 
sample $\xvec_i\in\Xmat$, we add an extra perturbation $\Delta\deltavec_i$ with minimal norm such as the perturbed point $\xvec_i+\deltavec+\Delta\deltavec_i$ fools the classifier. This is done by solving
\begin{align}
    \Delta\deltavec_i = \argmin_\rvec \left\|\rvec\right\|_{p} \mathrm{s.t.}\; g(\xvec_i+\deltavec+\rvec)\ne g(\xvec_i)
\end{align}
where $g$ is the classification network. Then, we enforce the constrain $\left\|\deltavec\right\|\le\varepsilon$, by projecting the updated perturbation on the $\ell_p$ ball of radius $\varepsilon$.
The iterations stop when at least a proportion $P$ of the samples in $\Xmat$ is miss-classified.

Moosavi et al.~\cite{moosavi2017universal} showed that universal perturbations present cross-model universality. In other words, 
universal perturbations computed for a specific architecture can also fool other architectures. Hence, it poses a potential security breach to break any classifier. Motivated by this, we used the universal perturbations pre-computed in Armory toolkit\footnote{\url{https://github.com/twosixlabs/armory/blob/master/armory/data/adversarial/librispeech_adversarial.py}} to create transfer black-box attacks for our x-vector classifier. These universal perturbations were create using a white-box SincNet classification model~\cite{ravanelli2018speaker}\footnote{\url{https://github.com/mravanelli/SincNet}}.


 \begin{figure*}
    \centering
    \includegraphics[width=0.7\textwidth]{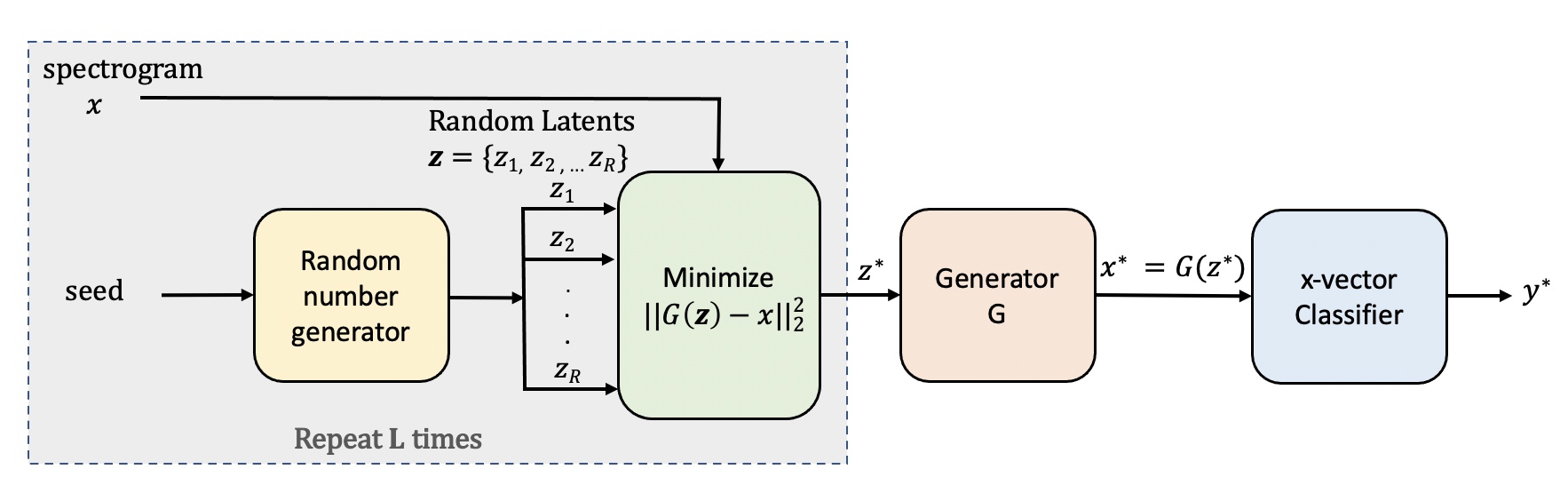}
    \caption{Scheme of Defense-GAN inference step}
    \label{fig:DefenseGAN}
    \vspace{-5mm}
\end{figure*}
\vspace{-3mm}
\section{Adversarial Defenses}
\label{sec:defenses}

\subsection{Adversarial Training}
\label{sec:advtrain}

The most simple and intuitive direction for training an adversarial robust model $\theta$ is to generate adversarial examples on-the-fly and utilize them  to train the model. This form of data augmentation is expected to cover areas of the input space not included in regular samples and improve adversarial robustness.  Adversarial training~\cite{madry2018towards} consists of a mini-max optimization given by
\begin{equation}
\underset{\boldsymbol{\theta}}{\operatorname{argmin}} \quad \mathbb{E}_{(\boldsymbol{x}, y^{\mathrm{benign}}) \sim \mathcal{D}}\left[\max _{\boldsymbol{\delta}:\|\boldsymbol{\delta}\|_{p}<\varepsilon} \quad L(\boldsymbol{x}+\boldsymbol{\delta}, y^{\mathrm{benign}}, \boldsymbol{\theta})\right].
\end{equation}
For the inner maximization operation, the PGD attack algorithm has proven to be a good defense~\cite{jati_adversarial_2020}. The outer minimization operation follows the standard training procedure.

However, robustness degrades when test attacks differ from those used
when training the system, e.g., if a different $\varepsilon$ or threat model or algorithm is used. In addition, adversarial training significantly increases training time, hindering scalability when using very large x-vector models. Another problem is the high cost of generating strong adversarial examples.

\subsection{Randomized smoothing} 
\label{sec:smooth}

Randomized smoothing is a method to construct a new \emph{smoothed} classifier $g'$ from a base classifier $g$. The smoothed classifier $g'$ predicts the 
class that the base classifier $g$ would predict given an input $\xvec$ perturbed with isotropic Gaussian noise~\cite{cohen2019certified}. That is
\begin{align}
    &g'(\xvec)=\argmax_c \mathbb{P}(g(\xvec+\nvec)=c)\\
    &\nvec \sim \Gausstwo{\zerovec}{\sigma^2\Imat}
\end{align}
where $\mathbb{P}$ is a probability measure and the noise standard deviation $\sigma$ is a hyperparameter that controls the trade-off
between robustness and accuracy. 
Cohen et al.~\cite{cohen2019certified} proved tight bounds for certified accuracies under Gaussian noise smoothing. Despite this defense
is not certifiable for $L_p$ with $p\ne2$, we found that it also performs well for other norms like $L_\infty$.

The main requirement for this defense to be effective is that the base classifier $g$ needs to be robust to Gaussian noise.
The base x-vector classifier is already robust to noise since it is trained on speech augmented with real noises~\cite{musan2015} and reverberation~\cite{ko2017study} using a wide range
of signal-to-noise ratios and room sizes. However, we robustified our classifier further by adding an extra fine-tuning step.
In this step, we fine-tune all the layers of the x-vector classifier by adding Gaussian noise on top of the real noises and reverberations. Instead of matching train and test $\sigma$ values, we sample train $\sigma$ from the uniform distribution in the interval $[0, 0.3]$, assuming a waveform with dynamic range [-1,1] interval. 

\begin{figure*}
    \centering
    \captionsetup{justification=centering}
    \includegraphics[width=0.7\textwidth]{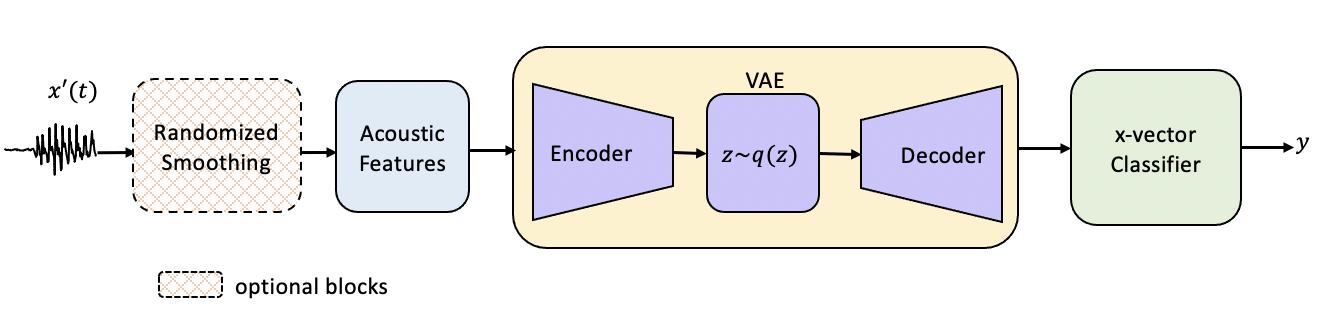}
    \caption{Pipeline of VAE defense. Combination of VAE with randomized smoothing is indicated by optional block}
    \label{fig:vae_pipeline}
    \vspace{-5mm}
\end{figure*} 

\vspace{-3mm}
\subsection{Defense-GAN}
\label{sec:defgan}

Generative adversarial networks (GAN)~\cite{goodfellow_generative_2014} are implicit generative models 
where two networks, termed Generator $G$ and Discriminator $D$, compete with each other 
The goal of $G$ is to generate samples $G(\zvec)$ that are similar to real data given a latent random variable $\zvec\sim\stdnormal$. Meanwhile, $D$ learns to discriminate between real and generated (fake) samples.
$D$ and $G$ are trained alternatively to optimize
\begin{multline}
\label{eq:gan_minimax}
  V(G,D) = \Expcond{\log (D(\xvec))}{\xvec\sim p_{\mathrm{data}}(\xvec)} \\ + \Expcond{\log(1-D(G(\zvec)))}{\zvec\sim p_z(\zvec)} \;,
\end{multline}
which is minimized w.r.t. $G$ and maximized w.r.t. $D$.
It can be shown that 
this objective minimizes the Jensen-Shannon (JS) distance between the real and generated distributions~\cite{goodfellow_generative_2014}.

Wasserstein GAN (WGAN) is an alternative formulation that minimizes the Wasserstein distance rather than the JS distance~\cite{arjovsky2017wasserstein}. Using
Kantorovich-Rubinstein duality, the objective that achieves this is
\begin{align}
    V(G,D)= \Expcond{D(\xvec)}{\xvec\sim p_{\mathrm{data}}(\xvec)}-\Expcond{D(G(\zvec))}{\zvec\sim p_z(\zvec)}
\end{align}
where the network $D$, termed as \emph{critic}, needs to be a K-Lipschitz function. Wasserstein GAN usually yields better performance than vanilla GAN.


Samangouei et al.~\cite{samangouei_defense-gan_2018} proposed to use GANs as a pre-processing defense against adversarial attacks for images. This method, termed Defense-GAN, has been demonstrated as an effective defense in several computer vision datasets. The key assumptions are that a well-trained generator learns the low dimensional manifold of the clean data while adversarial samples are outside this manifold. Therefore, projecting the test data onto the manifold defined by the GAN Generator will clean the sample of the adversarial perturbation.

Defense-GAN follows a two-step approach. In the training phase, a GAN is trained on benign data to learn the low-dimensional data manifold. In the inference phase, the test sample $\xvec$ is projected onto the GAN manifold by optimizing latent variable $\textbf{z}$ to minimize the reconstruction error
\begin{equation}
    \label{eq:defgan}
        \zvec^{*}=\argmin_{\textbf{z}} \left\| G(\textbf{z}) - \xvec \right\|^2_2\;.
\end{equation}
Then, the projected sample is $\xvec^{*}=G(\zvec^{*})$ where $\zvec$ is optimized by $L$ gradient descent iterations, initialized with $R$ random seed (random restarts). Thus $R\times L$ forward/backward passes are carried out on the Generator. Figure~\ref{fig:DefenseGAN} outlines the Defense-GAN inference step.


 We adapted DefenseGAN to the audio domain. We trained Wasserstein GAN with a Deep Convolutional generator-discriminator (DCGAN) to generate clean log-Mel-Spectrograms, similar to~\cite{donahue_adversarial_2019}. We enforced the 1-Lipschitz constraint for the Discriminator by adding a gradient penalty loss~\cite{gulrajaniimproved2017}. The Generator was trained to generate spectrogram patches of 80 frames per latent variable. At inference, we used the GAN to project adversarial spectrograms to the clean manifold. Note that adversarial perturbation was added in the wave domain, not in the log-spectrogram domain.
The DefenseGAN procedure needed to be sequentially applied to each block of 80 frames until the full utterance spectrogram is reconstructed. 
We observed that DefenseGAN produced over-smoothed spectrograms, which degraded performance on benign samples. 
To mitigate this, we used a weighted combination of original $\xvec$ and reconstructed
spectrograms $\xvec^{*}$ as $\xvec^{**}=\alpha\xvec^{*}+(1-\alpha)\xvec$.
Another parallel study~\cite{esmaeilpour2021} proposing DefenseGAN for defense against ASR adversarial attacks uses class embeddings for the GAN generator in addition to noise vector. However, they too encounter over-smoothed spectrograms and mode collapse issues. 
The main advantage of Defense-GAN is that it is not trained for any particular type of attack, so it is ``attack independent''. Another advantage is that the iterative procedure used for inference is not differentiable. Thus, it is difficult to create adaptive attacks, needing to resort to backward-pass differentiable approximation (BPDA)~\cite{athalye2018obfuscated} (see Section~\ref{sec:threat}). Assuming small adversarial perturbations, we approximate $\xvec'\approx\xvec$ for the backward pass for BPDA.

DefenseGAN also has some disadvantages. The success of Defense-GAN is highly dependent on training a good Generator~\cite{samangouei_defense-gan_2018},
otherwise, the performance suffers on both original and adversarial examples. Also, it a very slow procedure as it required $R\times L$ forward/backward passes on the Generator. Moreover, hyper-parameters $L$ and $R$ need to be tuned to balance speed and accuracy.


\vspace{-3mm}    
\subsection{Variational autoencoder defense}
\label{sec:vae}

Considering the shortcomings of DefenseGAN, we decided to explore other generative models such as
variational auto-encoders (VAE)~\cite{kingma-iclr14}.  Variational auto-encoders are probabilistic 
latent variable models consisting of two networks, termed encoder and decoder. The decoder defines the generation model of the observed variable $\xvec$ given the latent $\zvec$, i.e., the 
conditional data likelihood $\Prob{\xvec|\zvec}$. Meanwhile, the encoder computes the approximate posterior
of the latent given the data $q(\zvec|\xvec)$. The model is trained by maximizing the evidence lower bound (ELBO) given by
\begin{align}
    \lowb = \Expcond{\lnProb{\xvec|\zvec}}{q(\zvec|\xvec)} - \DKL{q(\zvec|\xvec)}{\Prob{\zvec}},
\end{align}
where $\Prob{\zvec}$ is the latent prior, typically standard normal. The first term of the ELBO intends to minimize
the reconstruction loss, while the second is a regularization term that keeps the posterior close to the prior. 
The KL divergence term creates an information bottleneck (IB), limiting the amount of information flowing through
the latent variable~\cite{alemi2016deep}. We expect this IB to pass the most relevant information while removing small adversarial perturbations from the signal.

\begin{figure*}
    \centering
    \captionsetup{justification=centering}
    \includegraphics[width=0.9\textwidth]{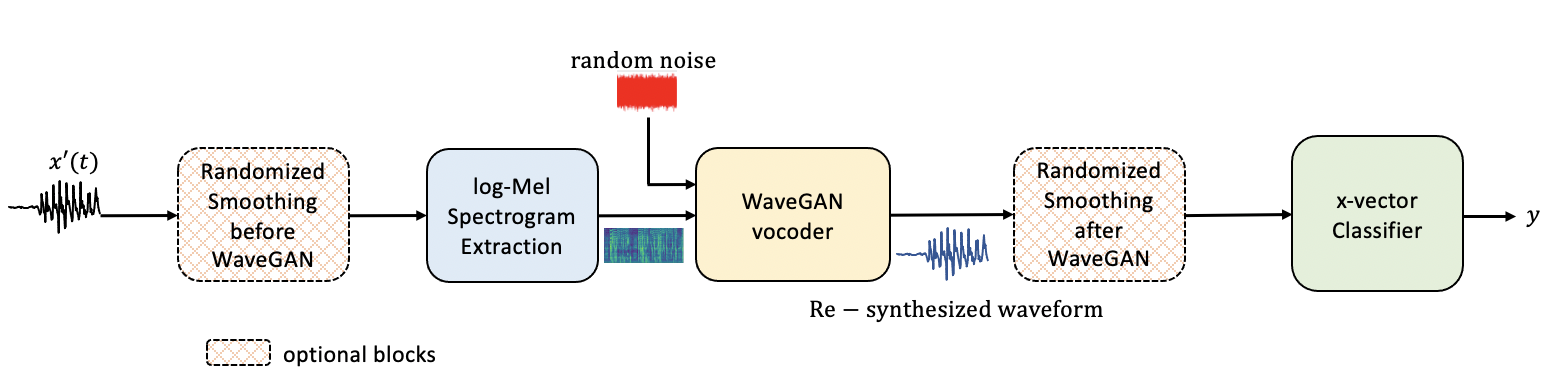}
    \caption{Pipeline for PWG Defense. Combination of PWG with randomized smoothing is indicated by the optional blocks. }
    \label{fig:pipelinewavegan}
    \vspace{-5mm}
\end{figure*}


As for DefenseGAN, we applied this defense in the log-filter-bank domain, as shown in Figure~\ref{fig:vae_pipeline}. 
We assumed Gaussian forms for the likelihood and the approximate posterior. The encoder and decoder predict the means and variances of these distributions.
To make our VAE more robust, we trained it in a denoising fashion~\cite{im2017denoising} (Denoising VAE). During training, the latent posterior is computed from a noisy version of the sample, while the decoder tries to predict the original sample.
At inference, we compute the latent posterior from the adversarial sample, draw a sample $\zvec\sim q(\zvec|\xvec)$; and forward pass $\zvec$ through the decoder. We used the decoder predicted mean as an estimate of the benign sample.
Variational auto-encoders have the advantage of being easier to train than GANs, and inference is also faster since
it consists of a single forward pass of encoder/decoder networks. Nevertheless, they have the drawback that they are fully differentiable so
and the adversary can adapt to the defense back-propagating gradients through the VAE. In that sense, we considered the two types of adaptive attacks described in Section~\ref{sec:threat}: attacks where the adversary doesn't know the parameters of VAE, so it needs to approximate the gradients by BPDA; and attacks where the adversary can compute the gradients exactly (termed E2ED).

We also experimented with combining Randomized Smoothing and VAE defenses, as illustrated in Figure~\ref{fig:vae_pipeline}.



\vspace{-3mm}
\subsection{ParallelWaveGAN (PWG) vocoder defense} 
\label{sec:wavegan}
For a defense very inherent to speech systems, we propose using a vocoder as a defense. A vocoder reconstructs the speech waveform given a compressed representation of it, in our case log-Mel-Spectrograms. We used the ParallelWaveGAN (PWG) vocoder proposed by Yamamoto et al.~\cite{yamamoto2020parallel}.  This vocoder is based on a generative adversarial network, similar to DefenseGAN. However, there are some differences
\begin{itemize}
    \item DefenseGAN reconstructs the clean spectrograms while PWG reconstructs the clean waveform.
    \item In DefenseGAN, reconstruction is based only on the latent variable $\zvec$. Meanwhile, in the PWG vocoder, both generator $G(\yvec,\zvec)$ and discriminator $D(\xvec,\yvec)$ are
    also conditioned on the log-Mel Spectrogram $\yvec$ of the input signal.
    \item At inference, DefenseGAN needs to iterate to find the optimum value of the latent $\zvec$. In the PWG, we just need to forward pass the log-Mel-Spectrogram $\yvec$ and the latent $\zvec$ through the PWG generator. $\zvec$ is a standard Normal random variable with the same length as the input waveform. Figure~\ref{fig:pipelinewavegan} shows the reconstruction pipeline.
\end{itemize}
PWG is trained on a combination of waveform-domain adversarial loss; and a reconstruction loss in Short Time Fourier Transform (STFT) domain at multiple resolutions. The reconstruction loss improves the stability and efficiency of the adversarial training. It is the sum of several STFT losses computed with different spectral analysis parameters (window length/shift, FFT length). 
The architecture of the generator is a non-autoregressive WaveNet, while the architecture for the 
Discriminator is based on a dilated convolutional network. As this defense can be back-propagated through, we considered both adaptive attack alternatives, BPDA and E2ED, similar to VAE defense. 


As shown by the optional blocks in Figure~\ref{fig:pipelinewavegan}, we can also combine smoothing and PWG defenses. Here, we can choose whether we want to apply smoothing before or after PWG.


    

\vspace{-3mm}
\section{Experimental Setup}
\label{sec:exp}
\subsection{Speaker identification dataset and task}
\label{sec:data}

The DARPA program \emph{Guaranteeing AI Robustness Against Deception} (GARD)\footnote{\url{https://www.darpa.mil/program/guaranteeing-ai-robustness-against-deception}}
promotes research on defenses against adversarial attacks in different domains. The program organizes periodic evaluations to benchmark the defenses developed by the participants.
Our experimental setup is based on the most recent GARD evaluation on defenses for speaker identification systems.
The proposed task is closed-set speaker identification, also known as speaker classification. Given a test utterance, the task is to identify the most likely among the enrolled speakers. 

The speaker identification task is built on the Librispeech~\cite{panayotov2015librispeech} development set, 
which contains 40 speakers (20 male and 20 female). 
Each speaker has about 4 minutes for enrollment and 2 minutes for test, with a total of 1283 test recordings.
While the setup is small, we need to consider the extensive computing cost of performing adversarial attacks.
For example, Carlini-Wagner attacks require hundreds of gradient descent iterations to obtain the adversarial perturbation.
Also, note that following Doddington's rule of 30~\cite{Doddington2000}, with 1283 recordings, we can measure error rates as low as 
2.3\% with confidence (97.7\% accuracy).
Thus, this task allowed us to obtain meaningful conclusions while limiting computing costs.

We will compare the performance of attacks and defenses using classification accuracy as the primary metric. Attacks will degrade the accuracy, while defenses will try to increase accuracy to match up to the performance of the non-attacked systems. To quantify the strength of the attack, we will indicate the $L_\infty$ norm of the adversarial perturbation, assuming that the waveform's dynamic range was normalized to $[-1,1]$.

This experimental setup is publicly available through the Armory toolkit\footnote{\url{https://github.com/twosixlabs/armory/blob/master/docs/scenarios.md}}. 
Armory toolkit uses the attacks available in the Adversarial Robustness Toolbox\footnote{\url{https://github.com/Trusted-AI/adversarial-robustness-toolbox}} to generate the adversarial samples.

\subsection{Speaker verification task}
\label{sec:data_sv}

We also evaluated our best defenses in a speaker verification task. In this case, we used VoxCeleb1 \emph{Original-Clean} trial list (37k trials). Note that these experiments have high computing costs (8 RTX 2048 GPUs were used), so it was not feasible to evaluate on the larger Entire and Hard lists. This is the same experimental setup we used in our previous work~\cite{villalba2020x}. The enrollment side of the trial was assumed to be benign and known to the adversary, while the adversary adds the adversarial perturbation to the test waveform.
We used cosine scoring to compare the enrollment and test x-vector embeddings. Scores were calibrated by linear logistic regression on the bonafide trials. Unlike in our previous work, we limited the duration of the test utterances to five seconds to fit the defense and x-vector models in the GPU simultaneously and perform adaptive attacks.

\vspace{-3mm}
\subsection{Baseline}
\label{sec:baseline}

We experimented with different x-vector architectures for our undefended baseline. Adversarial perturbations were added in the time domain, i.e., on the speech waveform. Acoustic features were 80 dimension log-Mel-filter-banks with short-time mean normalization. We used differentiable log-Mel-filter-banks implementation. These features were used as input
to x-vector networks. 
The network predicts logits for the speaker classes. 
The full pipeline was implemented in PyTorch~\cite{Paszke2019} so we can back-propagate gradients from the final score to the waveform. 
We evaluated architectures based on ResNet34, EfficientNet-b0/b4, Transformer-Encoder, ThinResNet34 and their fusion. For a fair comparison, we used mean and standard deviation over time for pooling for all architectures. The x-vector networks were pre-trained on the VoxCeleb2 dataset~\cite{Nagrani2020}, comprising 6114 speakers. Training data was augmented $6\times$ with noise from the MUSAN corpus\footnote{\url{http://www.openslr.org/resources/17}}
and impulse responses from the RIR dataset\footnote{\url{http://www.openslr.org/resources/28}}. We used
additive angular margin softmax loss (AAM-Softmax)~\cite{Deng2019} with margin=0.3 and scale=30, and Adam optimizer to train the network.
To use this network for speaker identification on the LibriSpeech dev set, we restarted the last two layers of the network
and fine-tuned them on the LibriSpeech dev train partition. We fine-tuned with AAM-Softmax with margin=0.9, since increasing the margin
slightly improved performance in this set.

\vspace{-3mm}
\subsection{Adversarial Attacks}

We experimented with attacks implemented in the Adversarial Robustness Toolkit: FGSM, BIM, PGD, CW-L2, and Universal Perturbations. For FGSM and BIM, we tested $L_\infty$ norms $\varepsilon$ between 0.0001 and 0.2. For BIM, we used a learning rate $\alpha=\varepsilon/5$, and performed 7, 50, and 100 iterations. For PGD, we used a learning rate $\alpha=\varepsilon/5$, 10 random restarts, and performed 50 and 100 iterations. Universal perturbations were transferred from a SincNet model~\cite{ravanelli2018speaker}, to create transfer black-box attacks as explained in Section~\ref{sec:universal}. For CW, we 
used confidence $\kappa=0$, learning rate 0.001, 10 iterations in the inner loop, and a maximum of 10 iterations in the outer loop.

\vspace{-3mm}
\subsection{Defenses}

\subsubsection{Adversarial training defense}
Adversarial training was our baseline defense. 
As in the baseline model, we initialized the model with the x-vector model pre-trained on VoxCeleb2. Then, we fine-tuned the last two layers of the model on the LibriSpeech speaker identification task to adapt to the new classes. Finally, we fine-tuned the full model with adversarial samples generated on-the-fly as described in Section~\ref{sec:advtrain}. We compared FGSM and PGD-$L_\infty$ algorithms to generate the attacks used for adversarial training. We also compared using a single value for $\varepsilon \in \unif[0.0001, 0.1]$ for the whole training or sampling a different $\varepsilon$ from a Uniform distribution in each model update iteration. For the PGD attacks, we used 10 iterations, and an optimization step $\alpha=\varepsilon/5$. This is similar to Jati et al.~\cite{jati_adversarial_2020}. However, our work considers much stronger (and perceptible) attacks till $L_{\infty}=0.1$ (much larger than maximum $L_{\infty}=0.005$ considered in~\cite{jati_adversarial_2020}) and proposes adversarial training using uniforming sampling instead of using a single $\varepsilon$.

\subsubsection{Randomized smoothing defense} 
To improve the performance for the randomized smoothing defense, we fine-tuned the network with Gaussian noise with $\sigma \sim \mathcal{U}(0.0.3)$ on top of real noises and reverberation. We compared three versions: fine-tuning the network classification head (last two layers) without Gaussian noise; fine-tuning the head network with Gaussian Noise, and fine-tuning the full network with Gaussian noise. At inference time, we used a single noise sample to create the smoothed classifier, since using more did not yield improvements.

\begin{table*}
\centering
\caption{\label{tab:ana_archi}
Classification accuracy (\%) for several undefended x-vector architectures under adversarial attacks}
\begin{tabular}{@{}lccccccccccccc@{}}
\toprule
\textbf{Architecture}  & \textbf{Clean}  & \multicolumn{5}{c}{\textbf{FGSM Attack}} & \multicolumn{5}{c}{\textbf{BIM Attack}} & \textbf{Universal} & \textbf{CW} \\
\cmidrule(r){1-1}\cmidrule(lr){2-2}\cmidrule(lr){3-7}\cmidrule(lr){8-12}\cmidrule(l){13-13}\cmidrule(l){14-14}
$L_{\infty}$ &  & 0.0001 & 0.001 & 0.01 & 0.1 & 0.2  & 0.0001 & 0.001 & 0.01 & 0.1 & 0.2 & 0.3 & - \\
\midrule

1. ResNet34 & 100.0  & 99.1  & 95.8  & 95.6  & 93.3  & 87.2  & 92.2  & 14.8  & 0.0  & 0.0  & 0.0  & 100.0  & 1.3 \\
2. EfficientNet-b0  & 100.0  & 99.2  & 95.6  & 93.0  & 93.6  & 88.1  & 96.9  & 27.7  & 0.0  & 0.0  & 0.0  & 100.0  & 0.8 \\
3. EfficientNet-b4  & 100.0  & 99.5  & 95.8  & 92.3  & 93.1  & 88.8  & 98.1  & 30.5  & 0.0  & 0.0  & 0.0  & 100.0  & 0.0 \\
4. Transformer  & 99.5  & 96.3  & 80.6  & 76.4  & 49.5  & 32.1  & 81.9  & 20.3  & 0.2  & 0.0  & 0.0  & 99.9  & 1.9 \\
5. ThinResNet34  & 100.0 & 98.0 & 91.1 & 89.2 & 85.6 & 74.5 & 88.0 & 2.2 & 0.0 & 0.0 & 0.0 & 100.0 & 1.1 \\
Fusion 2+4+5  & 100.0  & 99.8  & 97.5  & 97.0  & 88.4  & 78.0  & 98.9  & 66.4  & 16.1  & 0.0  & 0.2  & 100.0  & 49.1 \\
\bottomrule
\end{tabular}
\vspace{-1mm}
\end{table*}


\subsubsection{Defense-GAN}

The GAN in DefenseGAN used a Deep Convolutional generator-discriminator (DCGAN) trained with Wasserstein and gradient penalty losses, similar to Donahue et al.~\cite{donahue_adversarial_2019}. In addition, it was trained on VoxCeleb2 using the same log-Mel-filter-bank features required by the x-vector networks. The generator and discriminator architectures are shown in Table~\ref{tab:defgan_archi}, in the Appendix. 





\subsubsection{Variational auto-encoder defense}

A denoising VAE was trained on VoxCeleb2 augmented with noise and reverberation to model the manifold of benign log-Mel-Spectrograms.
Contrary to DefenseGAN, the VAE improved the reconstruction error when increasing the complexity of the encoder/decoder architectures. The architectures that performed the best were based on 2D residual networks. Table~\ref{tab:vae_encdec} describes the Encoder/Decoder architectures. The encoder downsamples the input spectrogram in time and frequency dimension multiple times to create an information bottleneck. The output of the encoder is the mean of the posterior $q(\zvec|\xvec)$, while the variance is kept as a trainable constant
Meanwhile, the decoder is a mirrored version of the encoder. It takes the latent sample $\zvec\sim q(\zvec|\xvec)$ and upsamples it to predict the mean of $\Prob{\xvec|\zvec}$. The variance is also a trainable constant. To perform the upsampling operation, we used
subpixel convolutions, since they provide better quality than transposed convolutions~\cite{shi2016real}.


\subsubsection{PWG vocoder defense} 
\label{sec:exp_wavegan}

We used publicly available PWG implementation\footnote{\url{https://github.com/kan-bayashi/ParallelWaveGAN}} and compared the robustness of two models. First, the public model trained on the Arctic Voices~\cite{kominek2004cmu} datasets. Second, we trained a model on VoxCeleb2~\cite{chung2018voxceleb2}. 
 


\section{Results}
\label{sec:results}
\subsection{Undefended baselines}
\label{sec:res_baseline}
We compared the vulnerability of ResNet34, EfficientNet, Transformer, and ThinResNet34 x-vector to adversarial attacks. 
Table~\ref{tab:ana_archi} shows classification accuracy for undefended baselines under FGSM, BIM, CW, and universal transfer attacks.
The systems are robust to the Universal transfer attack obtaining accuracy similar to the clean condition. 
This indicates that the adversary cannot transfer attacks from white-box models that are very different from the victim model verifying the claim by Abdullah et al. ~\cite{abdullahsok2020}.
FGSM produces adversarial attacks quickly, but the attacks are not very strong to fool the classifier. We only observed strong degradation for $L_{\infty}\ge 0.1$. However, for iterative attacks (BIM, CW), all systems failed even with imperceptible perturbations. The system using a fusion of the above networks gives light improvements (especially for BIM attacks with  $L_{\infty} \le 0.1$ and CW); however, the system is still vulnerable. In the following sections, we analyze the defenses only using the ThinResNet34 x-vector.
This is mainly motivated by the high computing cost of performing adversarial attacks (ThinResNet34 is around $16\times$ faster than the full ResNet34), and for the sake of simplicity.


\begin{table*}
\centering
\caption{\label{tab:results_advtrain}
Classification accuracy (\%) for Adversarial training with FGSM and PGD. $\unif(a,b)$ denotes uniform sampling in the range a to b.}
\begin{tabular}{@{}llccccccccccccc@{}}
\toprule
\textbf{Defense} & \textbf{AdvTraining}  & \textbf{Clean}  & \multicolumn{5}{c}{\textbf{FGSM Attack}} & \multicolumn{5}{c}{\textbf{BIM Attack}} & \textbf{Universal} & \textbf{CW}\\
\cmidrule(r){1-1}\cmidrule(lr){2-2}\cmidrule(lr){3-3}\cmidrule(lr){4-8}\cmidrule(lr){9-13}\cmidrule(l){14-14}\cmidrule(l){15-15}
$L_{\infty}$ & \textbf{$\varepsilon$} & & 0.0001 & 0.001 & 0.01 & 0.1 & 0.2  & 0.0001 & 0.001 & 0.01 & 0.1 & 0.2 & 0.3 & - \\
\midrule
No defense  &  -   & 100.0 & 96.9 & 90.0 & 92.3 & 93.4 & 91.1 & 83.4 & 2.3 & 0.0 & 0.0 & 0.0 & 100.0 &  1.4 \\
\midrule
PGD AdvTr & 0.0001 & 98.8 & 98.4 & 89.8 & 58.1 & 48.3 & 44.8 & 97.5 & 55.9 & 0.8 & 0.6 & 0.6 & 99.5 & 1.4 \\
 & 0.001 & 97.2 & 96.6 & 94.7 & 77.8 & 32.3 & 27.8 & 96.7 & 90.3 & 19.8 & 1.6 & 1.6 & 98.0 & 4.4\\
 & 0.01 & 70.9 & 70.8 & 68.4 & 55.8 & 25.2 & 19.1 & 71.1 & 66.9 & 39.2 & 8.3 & 8.0 & 84.9 & 32.5\\
 & $\unif(0, 0.01)$ & 75.5 & 76.4 & 75.3 & 59.8 & 25.0 & 18.1 & 75.8 & 72.7 & 39.4 & 8.9 & 8.4 & 87.9 & 30.3\\
 \midrule
FGSM AdvTr & 0.0001 & 98.8 & 97.5 & 88.1 & 63.7 & 53.9 & 47.0 & 97.0 & 45.9 & 0.3 & 0.3 & 0.2 & 99.6 & 1.6\\
 & 0.001 & 97.5 & 97.7 & 95.0 & 80.9 & 49.2 & 44.8 & 97.5 & 80.9 & 7.2 & 1.1 & 0.8 & 98.9 & 7.5\\
 & 0.01 & 71.9 & 70.8 & 80.3 & 95.8 & 74.7 & 50.8 & 68.8 & 55.5 & 30.6 & 25.8 & 23.8 & 93.8 & 33.1\\
 & 0.1 & 46.4 & 45.8 & 49.4 & 65.3 & 96.1 & 94.1 & 45.0 & 36.1 & 22.2 & 37.5 & 42.2 & 64.0 & 21.1\\
 & $\unif(0, 0.01)$ & 89.1 & 89.2 & 88.3 & 89.5 & 63.6 & 49.8 & 89.1 & 77.0 & 24.5 & 7.2 & 7.0 & 95.9 & 32.3\\
 & $\unif(0, 0.1)$ & 72.8 & 70.8 & 78.1 & 92.0 & 94.1 & 91.3 & 69.8 & 50.9 & 22.0 & 20.9 & 21.9 & 90.2 & 23.1\\
\bottomrule
\end{tabular}
\end{table*}

\begin{table*}
\centering
\caption{\label{tab:results_smoothing}
Classification accuracy (\%) for randomized smoothing defense. We compare fine-tuning x-vector w/o Gaussian noise; and 
finetuning the network clasification head versus the full network.}
\begin{tabular}{@{}llccccccccccccc@{}}
\toprule
\textbf{Gaussian}  & \textbf{Smoothing}  & \textbf{Clean}  & \multicolumn{5}{c}{\textbf{FGSM Attack}} & \multicolumn{5}{c}{\textbf{BIM Attack}} & \textbf{Universal} & \textbf{CW}\\
\textbf{Augment}  & \textbf{$\sigma$}  \\
\cmidrule(r){1-1}\cmidrule(r){2-2}\cmidrule(lr){3-3}\cmidrule(lr){4-8}\cmidrule(lr){9-13}\cmidrule(l){14-14}\cmidrule(l){15-15}
$L_{\infty}$ &  & & 0.0001 & 0.001 & 0.01 & 0.1 & 0.2  & 0.0001 & 0.001 & 0.01 & 0.1 & 0.2 & 0.3 & - \\
\midrule
No & 0 & 100.0 & 98.0 & 91.1 & 89.2 & 85.6 & 74.5 & 88.0 & 2.2 & 0.0 & 0.0 & 0.0 & 100.0 & 1.1 \\
& 0.01 & 100.0 & 100.0 & 100.0 & 67.8 & 71.6 & 64.1 & 99.8 & 99.4 & 0.0 & 0.2 & 0.0 & 99.9 & 7.5 \\
& 0.1 & 98.0 & 98.0 & 97.5 & 94.4 & 26.4 & 25.8 & 97.8 & 97.5 & 90.3 & 2.5 & 2.2 & 98.7 & 83.4  \\
& 0.2 & 87.3 & 88.8 & 87.3 & 85.8 & 30.2 & 19.5 & 86.4 & 85.3 & 82.2 & 9.5 & 6.7 & 90.3 &  87.3\\
\midrule
Yes & 0 & 100.0 & 96.9 & 90.0 & 92.3 & 93.4 & 91.1 & 83.4 & 2.3 & 0.0 & 0.0 & 0.0 & 100.0 & 1.4 \\
Head  & 0.01 & 100.0 & 100.0 & 99.2 & 78.8 & 83.1 & 81.6 & 100.0 & 99.1 & 0.6 & 0.0 & 0.0 & 100.0 & 8.8 \\
& 0.1 & 98.6 & 98.6 & 98.9 & 97.8 & 47.3 & 45.3 & 99.2 & 98.6 & 97.3 & 0.9 & 1.1 & 99.5 & 81.6 \\
& 0.2 & 97.0 & 97.5 & 96.7 & 96.7 & 51.9 & 35.8 & 97.7 & 97.2 & 95.8 & 7.5 & 1.7 & 98.5 & 95.3 \\
\midrule
Yes & 0 & 99.4 & 97.0 & 90.5 & 94.5 & 94.8 & 90.6 & 88.0 & 17.2 & 0.5 & 0.5 & 0.3 & 99.9 & 2.0\\
Full-Net & 0.2 & 98.0 & 98.3 & 98.4 & 97.0 & 64.4 & 44.1 & 97.2 & 97.8 & 97.7 & 18.9 & 2.0 & 98.7 & 96.9\\
& 0.4 & 92.5 & 92.2 & 94.5 & 93.1 & 65.3 & 40.9 & 93.1 & 93.1 & 91.9 & 44.8 & 9.5 & 95.3 & 93.1\\
\bottomrule
\end{tabular}
\end{table*}

\subsection{Adversarial training defense}
Table \ref{tab:results_advtrain}  shows  the  results  for  adversarial training. We experimented with FGSM and PGD adversarial training with different training $\varepsilon$, i.e., $L_{\infty}$. Matching the adversarial training algorithm and the attack performed better, as expected. That is, FGSM adversarial training performed better for FGSM while PGD adversarial training performed better for BIM (a.k.a. PGD-$L_\infty$.) However, for FGSM attacks, the undefended system still was better. For training $\varepsilon=0.001$, we were able to protect the system for BIM $\varepsilon \le 0.001$. Increasing training $\varepsilon$ we could increase the protection for larger $\varepsilon$ and Carlini-Wagner. However, this implies an unacceptable deterioration of the benign performance. If the training $\varepsilon$ is too high--e.g., FGSM adversarial training with $\varepsilon=0.1$--the model over-fits to work on adversarial attacks, and adversarial performance becomes superior to benign performance. In the case of PGD, adversarial training with $\varepsilon > 0.01$ was not able to converge. We also tried to sample the value of the training $\varepsilon$ uniformly. In this case, adversarial performance was close to the case of doing adversarial training with the maximum $\varepsilon$ value, while benign performance was a little better.


\vspace{-3mm}
\subsection{Randomized smoothing defense}

Table~\ref{tab:results_smoothing} shows the results for the randomized smoothing defense.
We experimented with different Gaussian noise standard deviations $\sigma$. 
The first block of the table shows results for ThinResNet34 fine-tuned without Gaussian noise (only real noise and reverberation).
Smoothing only damages clean and universal attack performance when using very high $\sigma=0.2$, which is very perceptible noise.
We also observe that smoothing damaged accuracy under FGSM attacks as we increase $\sigma$. We hypothesize that the combination of adversarial and Gaussian noise is too loud for the speaker recognition system to perform well. 
Note that $L_\infty=0.1$ is very perceptible noise compared to the values of the speech samples, which follow a Laplacian distribution and, therefore, samples are concentrated around zero.
However, increasing $\sigma$ improved accuracy under BIM and CW attacks. To make the system robust to BIM attacks, we needed $\sigma\sim 10\times L_{\infty}$.
It also provided a huge improvement for the Carlini-Wagner attack w.r.t. the undefended system. 

In the second block, we fine-tuned the network head with Gaussian noise on top of real noise and reverberation. By doing this, 
we increased the robustness of the system for the smoothing defense. Thus, the performance for clean and universal with high smoothing $\sigma$ was improved, being close to the performance without defense. It also significantly improved the accuracy under FGSM with $L_\infty\in [0.01, 0.2]$--in some cases up to 100\% of relative improvement w.r.t the system fine-tuned without Gaussian noise. It also improved BIM at $L_{\infty}=0.01$. 

In the third block, we fine-tuned the full x-vector network with Gaussian noise. This allowed us to obtain good accuracies for
very high smoothing $\sigma\in[0.2,0.4]$. In most attacks, we got similar or better results than when fine-tuning just the head. For $\sigma=0.2$, we obtained the best results overall. For $\sigma=0.4$, we achieved the best results under BIM attack with $L_{\infty}\ge0.1$, at the cost of damaging accuracy in low $L_{\infty}$ attacks.
 

\begin{table*}
\centering
\caption{\label{tab:results_defgan}
Classification accuracy (\%) for DefenseGAN defense with $\alpha = 0.5$}
\begin{tabular}{@{}lccccccccccccc@{}}
\toprule
\textbf{DefenseGAN}  & \textbf{Clean}  & \multicolumn{5}{c}{\textbf{FGSM Attack}} & \multicolumn{5}{c}{\textbf{BIM Attack}} & \textbf{Universal} & \textbf{CW}\\
\cmidrule(r){1-1}\cmidrule(lr){2-2}\cmidrule(lr){3-7}\cmidrule(lr){8-12}\cmidrule(l){13-13}\cmidrule(l){14-14}
$L_{\infty}$ &  & 0.0001 & 0.001 & 0.01 & 0.1 & 0.2  & 0.0001 & 0.001 & 0.01 & 0.1 & 0.2 & 0.3 & - \\
\midrule
No Defense  & 100.0 & 96.9 & 90.0 & 92.3 & 93.4 & 91.1 & 83.4 & 2.3 & 0.0 & 0.0 & 0.0 & 100.0 &  1.4 \\
\midrule
$\alpha = 0.5$  & 95.9 & 91.6 & 84.2 & 81.9 & 49.4 & 23.3 & 85.0 & 23.6 & 2.8 & 1.4 & 1.6 & 96.9 &  60.9 \\
\bottomrule
\end{tabular}
\end{table*}

\begin{table*}
\centering
\caption{\label{tab:results_vae}
Classification accuracy (\%) for VAE Defense with and without randomized smoothing; with and without fine-tuning the x-vector classifier with the VAE pre-processing; considering adaptive attacks on VAE approximating gradients (BPDA) or white-box end-to-end differentiable (E2ED). Smoothing is applied before VAE}
\resizebox{\textwidth}{!}{
\begin{tabular}{@{}lllccccccccccccc@{}}
\toprule
\textbf{x-Vector} & \textbf{Smooth} & \textbf{Adaptive} & \textbf{Clean}  & \multicolumn{5}{c}{\textbf{FGSM Attack}} & \multicolumn{5}{c}{\textbf{BIM Attack}} & \textbf{Universal} & \textbf{CW}\\
\textbf{fine-tuning}  & \textbf{$\sigma$} & \textbf{Attack Type} \\
\cmidrule(r){1-1}\cmidrule(lr){2-2}\cmidrule(lr){3-3}\cmidrule(lr){4-4}\cmidrule(lr){5-9}\cmidrule(lr){10-14}\cmidrule(l){15-15}\cmidrule(l){16-16}
$L_{\infty}$ &  & & & 0.0001 & 0.001 & 0.01 & 0.1 & 0.2  & 0.0001 & 0.001 & 0.01 & 0.1 & 0.2 & 0.3 & - \\
\midrule
No defense  &  -  &  -  & 100.0 & 96.9 & 90.0 & 92.3 & 93.4 & 91.1 & 83.4 & 2.3 & 0.0 & 0.0 & 0.0 & 100.0 &  1.4 \\
\midrule
No  & 0.0 &  \textit{BPDA}  & 90.8 & 91.6 & 88.8 & 77.8 & 51.4 & 44.1 & 89.4 & 71.4 & 34.2 & 4.4 & 2.5 & 91.8 &  71.1 \\
\cmidrule(l){2-16}
 & 0.0 &  \textit{E2ED}  & 91.6 & 86.6 & 75.5 & 63.8 & 45.3 & 38.4 & 78.8 & 22.3 & 4.1 & 1.9 & 0.9 & 91.8 &  52.5 \\
\midrule
Yes  & 0.0 &  \textit{BPDA}  & 98.9 & 98.4 & 98.1 & 94.2 & 91.4 & 84.8 & 94.7 & 67.2 & 12.7 & 0.9 & 1.1 & 99.9 &  56.1 \\
 & 0.1 &  \textit{BPDA}  & 96.9 & 96.9 & 97.0 & 96.9 & 75.2 & 57.5 & 96.9 & 97.5 & 96.6 & 25.6 & 1.4 & 98.7 &  98.0 \\
 & 0.2 &  \textit{BPDA}  & 95.2 & 95.8 & 95.5 & 94.7 & 79.5 & 51.4 & 96.6 & 95.2 & 94.5 & 63.1 & 9.8 & 97.5 &  95.5 \\
\cmidrule(l){2-16}
 & 0.0 &  \textit{E2ED}  & 99.4 & 96.9 & 94.1 & 94.2 & 91.4 & 84.8 & 92.3 & 35.3 & 1.3 & 0.9 & 0.5 & 99.8 &  50.2 \\
 & 0.1 &  \textit{E2ED}  & 96.6 & 97.0 & 96.4 & 95.8 & 64.5 & 53.3 & 97.5 & 97.8 & 94.7 & 2.2 & 0.9 & 98.5 &  96.9 \\
 & 0.2 &  \textit{E2ED}  & 95.2 & 95.8 & 96.1 & 95.2 & 65.6 & 50.0 & 96.1 & 95.8 & 94.2 & 19.1 & 2.7 & 97.4 &  95.8 \\
 \bottomrule
\end{tabular}}
\end{table*}

\vspace{-3mm}
\subsection{Defense GAN} 

As explained in Section~\ref{sec:defgan}, DefenseGAN generated over-smoothed spectrograms, hence degrading the benign performance.
To solve this, we interpolated the original and reconstructed spectrograms using a factor $\alpha$. 
Lower $\alpha=0.25$ did not improved w.r.t. the undefended systems. Higher $\alpha=0.75$, degraded benign accuracy down to 72\%. Thus,
we selected $\alpha=0.5$ to maintain high benign accuracy while improving robustness against BIM and CW attacks. Table~\ref{tab:results_defgan} compares DefenseGAN versus 
undefended classification accuracies.
DefenseGAN degraded FGSM attacks but improved BIM and CW attacks. For CW, 
 we got an accuracy of 60.94\% against 1.41\% without defense. This shows that DefenseGAN has good potential to defend against CW attacks. This finding is similar to DefenseGAN for CW attack on images~\cite{samangouei_defense-gan_2018}.

\begin{table*}
\centering
\caption{\label{tab:results_wavegan_smooth}
Classification accuracy (\%) for PWG defense with and without randomized smoothing; considering PWG as known (adaptive) or unknown (non-adaptive) to the adversary. PWG was trained on VoxCeleb2, except the pre-trained model, which was trained on Arctic. }
\begin{tabular}{@{}llccccccccccccc@{}}
\toprule
\textbf{Defense} & \textbf{RS$\sigma$}  & \textbf{Clean}  & \multicolumn{5}{c}{\textbf{FGSM Attack}} & \multicolumn{5}{c}{\textbf{BIM Attack}} & \textbf{Universal} & \textbf{CW}\\
\cmidrule(r){1-1}\cmidrule(lr){2-2}\cmidrule(lr){3-3}\cmidrule(lr){4-8}\cmidrule(lr){9-13}\cmidrule(l){14-14}\cmidrule(l){15-15}
$L_{\infty}$ &  & & 0.0001 & 0.001 & 0.01 & 0.1 & 0.2  & 0.0001 & 0.001 & 0.01 & 0.1 & 0.2 & 0.3 & - \\
\midrule
No defense  &   -  & 100.0 & 96.9 & 90.0 & 92.3 & 93.4 & 91.1 & 83.4 & 2.3 & 0.0 & 0.0 & 0.0 & 100.0 &  1.4 \\
 \midrule
Pretrained \textit{BPDA}& - & 96.4 & 97.3 & 97.8 & 90.3 & 33.9 & 12.7 & 96.6 & 98.3 & 93.8 & 64.4 & 36.3 & 97.6 &  95.8 \\
 \midrule
\textit{BPDA} & - & 99.5 & 99.5 & 99.7 & 99.1 & 86.6 & 77.2 & 99.4 & 99.7 & 99.5 & 97.2 & 92.3 & 99.9 &  98.8 \\
\textit{E2ED} & -  & 97.0 & 98.8 & 95.8 & 93.6 & 83.0 & 62.3 & 94.7 & 36.4 & 0.8 & 0.8 & 0.8 & 99.8 &  37.5\\
 \midrule
Smoothing after PWG & 0.0 & 99.5 & 99.7 & 99.8 & 98.6 & 89.7 & 76.7 & 99.4 & 99.5 & 99.5 & 97.3 & 92.5 & 99.8 &  99.7 \\
\textit{BPDA}    & 0.001 & 99.7 & 99.5 & 99.7 & 99.2 & 85.0 & 71.7 & 99.7 & 99.7 & 99.4 & 97.0 & 92.3 & 99.9 &  99.5 \\
  & 0.01 & 99.6 & 99.4 & 99.5 & 99.7 & 87.5 & 64.8 & 99.5 & 99.7 & 99.7 & 97.2 & 92.7 & 99.9 &  99.5 \\
  & 0.1 & 97.1 & 97.3 & 97.2 & 96.4 & 89.1 & 65.3 & 97.3 & 97.0 & 96.4 & 95.8 & 90.5 & 98.6 &  97.0 \\
  & 0.2 & 91.1 & 91.1 & 91.4 & 91.4 & 85.3 & 62.0 & 92.3 & 91.3 & 90.9 & 89.2 & 85.9 & 94.0 &  91.3 \\
 \midrule
Smoothing before PWG & 0.0 & 99.5 & 99.5 & 99.7 & 99.1 & 86.6 & 77.2 & 99.4 & 99.7 & 99.5 & 97.2 & 92.3 & 99.9 &  98.8 \\
\textit{BPDA}  & 0.001 & 99.7 & 99.7 & 99.7 & 99.2 & 86.3 & 69.1 & 99.7 & 99.7 & 99.5 & 97.2 & 90.5 & 99.9 &  99.2 \\
  & 0.01 & 99.6 & 99.4 & 99.5 & 99.4 & 88.1 & 65.5 & 99.5 & 99.5 & 99.5 & 96.4 & 91.4 & 99.9 &  98.8 \\
  & 0.1 & 97.1 & 97.5 & 98.1 & 98.4 & 94.2 & 76.6 & 98.4 & 98.8 & 98.0 & 97.0 & 93.1 & 99.3 &  97.3 \\
   & 0.2 & 95.6 & 95.2 & 96.3 & 95.8 & 93.0 & 74.2 & 94.8 & 94.8 & 96.9 & 95.5 & 93.4 & 97.5 &  95.2 \\
   \cmidrule(l){2-15}
\textit{E2ED}    & 0.2 & 95.8 & 94.7 & 95.6 & 94.4 & 88.9 & 60.6 & 95.2 & 93.3 & 86.7 & 14.4 & 3.1 & 97.4 &  92.8\\
\bottomrule
\end{tabular}
\end{table*}

\begin{table*}
\centering
\caption{\label{tab:results_compare}
Summary of classification accuracy (\%) of all defenses with their best setting. Note: Smoothing $\sigma=0.2$,  PGD/FGSM AdvTr $\varepsilon= \unif(0, 0.01)$, PWG models is trained on Voxceleb. For adaptive attacks, PWG/VAE defenses are either approximated (BPDA) or end-to-end differentiable (E2ED) }
\begin{tabular}{@{}lccccccccccccc@{}}
\toprule
\textbf{Defense}  & \textbf{Clean}  & \multicolumn{5}{c}{\textbf{FGSM Attack}} & \multicolumn{5}{c}{\textbf{BIM Attack}} & \textbf{Universal} & \textbf{CW}\\
\cmidrule(r){1-1}\cmidrule(lr){2-2}\cmidrule(lr){3-7}\cmidrule(lr){8-12}\cmidrule(l){13-13}\cmidrule(l){14-14}
$L_{\infty}$ & -  & 0.0001 & 0.001 & 0.01 & 0.1 & 0.2  & 0.0001 & 0.001 & 0.01 & 0.1 & 0.2 & 0.3 & - \\
\midrule
No defense  & 100.0 & 96.9 & 90.0 & 92.3 & 93.4 & 91.1 & 83.4 & 2.3 & 0.0 & 0.0 & 0.0 & 100.0 &  1.4 \\
\midrule
PGD AdvTr  & 75.5 & 76.4 & 75.3 & 59.8 & 25.0 & 18.1 & 75.8 & 72.7 & 39.4 & 8.9 & 8.4 & 87.9 & 30.3\\
FGSM AdvTr & 89.1 & 89.2 & 88.3 & 89.5 & 63.6 & 49.8 & 89.1 & 77.0 & 24.5 & 7.2 & 7.0 & 95.9 & 32.3\\
\midrule
Smoothing   & 98.0 & 98.3 & 98.4 & 97.0 & 64.4 & 44.1 & 97.2 & 97.8 & 97.7 & 18.9 & 2.0 & 98.7 &  96.9 \\
\midrule
DefenseGAN  & 96.3 & 91.6 & 84.2 & 81.9 & 49.4 & 23.3 & 85.0 & 23.6 & 2.8 & 1.4 & 1.6 & 96.9 &  60.9 \\
\midrule
VAE \textit{BPDA}  & 98.9 & 98.4 & 98.1 & 94.2 & 91.4 & 84.8 & 94.7 & 67.2 & 12.7 & 0.9 & 1.1 & 99.9 &  56.1 \\
VAE \textit{E2ED}  & 99.4 & 96.9 & 94.1 & 94.2 & 91.4 & 84.8 & 92.3 & 35.3 & 1.3 & 0.9 & 0.5 & 99.8 &  50.2 \\
\midrule
Smoothing before VAE \textit{BPDA} & 95.2 & 95.8 & 95.5 & 94.7 & 79.5 & 51.4 & 96.6 & 95.2 & 94.5 & 63.1 & 9.8 & 97.5 &  95.5 \\
Smoothing before VAE \textit{E2ED}  & 95.2 & 95.8 & 96.1 & 95.2 & 65.6 & 50.0 & 96.1 & 95.8 & 94.2 & 19.1 & 2.7 & 97.4 &  95.8 \\
\midrule
PWG \textit{BPDA}  & 99.5 & 99.5 & 99.7 & 99.1 & 86.6 & 77.2 & 99.4 & 99.7 & 99.5 & 97.2 & 92.3 & 99.9 &  98.8 \\
PWG \textit{E2ED}  & 97.0 & 98.8 & 95.8 & 93.6 & 83.0 & 62.3 & 94.7 & 36.4 & 0.8 & 0.8 & 0.8 & 99.8 &  37.5 \\
\midrule
Smoothing before PWG \textit{BPDA}  & 95.6 & 95.2 & 96.3 & 95.8 & 93.0 & 74.2 & 94.8 & 94.8 & 96.9 & 95.5 & 93.4 & 97.5 &  95.2 \\
Smoothing before PWG \textit{E2ED}  & 95.8 & 94.7 & 95.6 & 94.4 & 88.9 & 60.6 & 95.2 & 93.3 & 86.7 & 14.4 & 3.1 & 97.4 &  92.8\\
\bottomrule
\end{tabular}
\end{table*}

\begin{table}
  \renewcommand{\arraystretch}{1}
    \centering
    \caption{\label{tab:WaveGANdefense_strong_attacks} Classification accuracy (\%) for PWG defense against strong white-box adaptive BIM and PGD. BIM attacks are for 50 or 100 iterations indicated by BIM-50/BIM-100. PGD attacks have 10 random restarts and  50 iterations (PGD-$50$) or 100 iterations (PGD-$50$), for $L_\infty = 0.001, 0.01$. Note: Smoothing $\sigma=0.2$}
    \vspace{-2mm}
    \setlength\tabcolsep{2pt}
    \begin{tabular}{@{}lcccccccc@{}}
        \toprule
        \textbf{Defense}  & \multicolumn{2}{c}{\textbf{BIM-50}}  & \multicolumn{2}{c}{\textbf{BIM-100}} & \multicolumn{2}{c}{\textbf{PGD-50}}  & \multicolumn{2}{c}{\textbf{PGD-100}}\\
        \cmidrule(r){1-1}\cmidrule(l){2-3}\cmidrule(l){4-5}\cmidrule(l){6-7}\cmidrule(l){8-9}
        $L_{\infty}$ & 0.001 & 0.01 & 0.001 & 0.01 & 0.001 & 0.01 & 0.001 & 0.01 \\
        \midrule
        No defense & 0.0 & 0.0 & 0.0 & 0.0 & 0.0 & 0.0 & 0.0 & 0.0 \\
        PWG & 99.5 & 99.4 & 99.5 &	99.4 & 99.4 & 99.7 & 99.8 & 99.5 \\
        Smoothing before & 95.6	& 96.3 & 94.4 & 95.8 & 95.6 & 96.7 & 96.1	& 96.3 \\
        PWG BPDA \\
        \bottomrule
    \end{tabular}
    \vspace{-4mm}
\end{table}

\begin{figure*}[h]
    \centering
    \captionsetup{justification=centering}
    \includegraphics[width=0.7\textwidth]{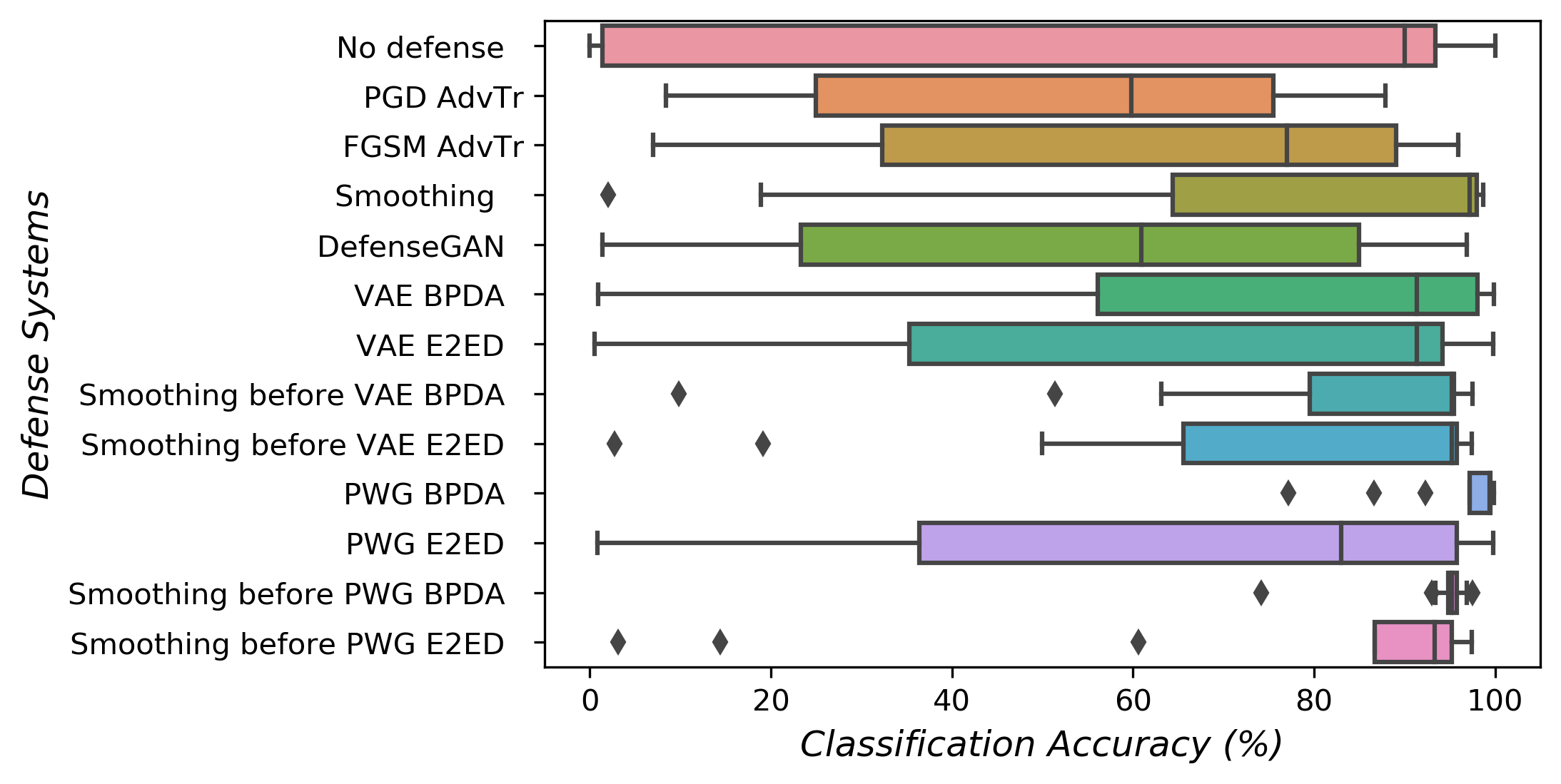}
    \caption{Summary of all defense systems with their best settings for all attack settings as in Table \ref{tab:results_compare} using boxplot}
    \label{fig:results_boxplot}
    \vspace{-4mm}
\end{figure*}

\subsection{Variational Auto-encoder (VAE) defense} 

Table~\ref{tab:results_vae} presents the VAE defense results. First, we used VAE as a pre-processing defense for the original ThinResNet34 x-vector (\emph{No fine-tuning} block in the table). We compared the case where the adversary does not have access to the fully differentiable VAE model and, hence, creates adaptive attacks using approximated gradients (BPDA); versus the case where the adversary has access to E2ED VAE. VAE BPDA degrades FGSM w.r.t. undefended for high $L_\infty>0.01$, which are perceptible. However, it provides significant improvements for BIM $L_\infty\le 0.01$--2.3\% to 71.4\% for $L_\infty=0.01$-- and Carlini-Wagner--1.4\% to 71.1\%. When using E2ED VAE, adversarial accuracies reduced. We only observe significant improvements w.r.t. undefended systems for BIM $L_\infty=0.01$ and Carlini-Wagner.
Introducing VAE degraded clean accuracy about 10\% absolute. To counteract this, we fine-tuned the x-vector model, including the VAE in the pipeline. Fine-tuning improved clean, FGSM, and universal transfer attack accuracies, being close to the ones
of the undefended system. Again, VAE improved BIM, and CW attacks w.r.t. the undefended system. Finally, we combined VAE and randomized smoothing defenses. Smoothing $\sigma=0.2$ for BPDA VAE improved all BIM and CW attacks with respect to just VAE. It also improved w.r.t. just smoothing $\sigma=0.2$ in Table~\ref{tab:results_smoothing}. Only BIM with $L_\infty=0.2$, which is very perceptible, was seriously degraded w.r.t to the clean condition. For the case of smoothing before E2ED VAE, BIM and CW also improved with respect to just E2ED VAE. However, there was no significant improvement w.r.t. just smoothing.



\subsection{PWG vocoder defense} 
\subsubsection{PWG defense on speaker identification}

Table~\ref{tab:results_wavegan_smooth} displays the PWG defense results. 
We started using PWG non-adaptive attacks and compared the publicly available model trained on Arctic to our model trained on the VoxCeleb test set. 
We observed that the VoxCeleb model performed better in the benign condition and across all attacks.
We conclude that the larger variability in terms of speakers and channels in VoxCeleb contributed to making the defense more robust.
Contrary to the VAE experiments above, PWG did not degrade the benign accuracy. 
Thus, we did not need to fine-tune the x-vector network, including the PWG in the pipeline. PWG provided very high adversarial robustness with accuracies $>90\%$ for most BPDA adaptive attacks. The average absolute improvement overall attacks
was 41.5\% w.r.t. the baseline. The improvement for BIM (80.47\% absolute) and CW (97.34\% absolute) is noteworthy. 
This is very superior to previous smoothing and 
VAE defenses. We also evaluated the robustness of PWG defense against strong PGD and BIM attacks (Table ~\ref{tab:WaveGANdefense_strong_attacks}) and show that PWG defense improves accuracy by $>97\%$ absolute on average for attacks with high number of iterations (50 and 100 iterations, $L_\infty$ of 0.001 and 0.01). However, PWG performance degraded significantly when the defense was made end-to-end differentiable. For E2ED adaptive attacks, PWG only improved in BIM $L_\infty <0.01$ and CW, similarly to E2ED VAE.


\begin{table*}
\centering
\caption{\label{tab:results_spkver}
Summary of Equal Error Rate (EER) for Speaker Verfication. The PWG model is trained on Voxceleb, and smoothing is applied before PWG.}
\begin{tabular}{@{}lcccccccccc@{}}
\toprule
\textbf{Defense}  & RSmooth & \textbf{Clean}  & \multicolumn{3}{c}{\textbf{FGSM Attack}} & \multicolumn{3}{c}{\textbf{BIM Attack}}  & \textbf{CW}\\
\cmidrule(r){1-1}\cmidrule(lr){2-2}\cmidrule(lr){3-3}\cmidrule(lr){4-6}\cmidrule(lr){7-9}\cmidrule(lr){10-10}
$L_{\infty}$ & $\sigma$  & -  & 0.0001 & 0.001 & 0.01 & 0.0001 & 0.001 & 0.01 &  lr=0.001  \\
\midrule
No defense & - & 2.20 & 18.73 & 37.96 & 38.22 & 48.89 & 50.00 & 50.00 & 50.00 \\
\midrule
PWG & 0 & 3.08 & 11.54 & 31.46 & 36.07 & 23.98 & 50.00 & 50.00 & 41.97 \\
PWG & 0.1 & 21.29 & 22.02 & 23.08 & 26.68 & 22.28 & 27.89 & 49.48 & 21.23 \\
\bottomrule
\end{tabular}
\end{table*}

We tried to improve PWG defense further by integrating smoothing with the PWG defense. 
We can do it in two ways: applying smoothing before or after the vocoder. The results are shown in the third and fourth blocks of Table~\ref{tab:results_wavegan_smooth}. We did not observe significant statistical differences between both variants for smoothing $\sigma<0.1$. For $\sigma\ge 0.1$, smoothing followed by PWG was better. However, overall, adding smoothing to the BPDA PWG did not provide consistent improvements w.r.t. just the PWG. The last line of the table shows the results for smoothing, followed by the PWG for the case when the adversary has access to E2ED PWG. In this case, accuracy
was similar to the case of just using smoothing. . Thus, we conclude that combining smoothing and PWG vocoder defense is the best strategy. When the adversary does not have access to the E2ED PWG defense model, it is robust across attacks and $L_\infty$ levels. However, when the adversary has an E2ED PWG model, it still maintains competitive robustness for most attacks, except those with very high $L_\infty$ norms. 


\subsubsection{PWG defense on speaker verification}

Finally, we evaluated our best defenses overall, i.e., PWG and Smoothing before PWD, 
on the open-set speaker verification task described in Section~\ref{sec:data_sv}.
We used the fully differentiable E2ED PWG vocoder for this experiment.
We limit our analysis to the best defenses due to the higher computing cost of 
speaker verification attacks w.r.t. identification attacks. This is because the 
number of attack evaluations is multiplied by the number of enrollment segments, while
for speaker identification, there is only one attack per test segment.
Table~\ref{tab:results_spkver} shows the results in terms of equal-error-rate EER(\%)
for FGSM, BIM, and CW-L2 attacks. 
For the undefended system, the table shows a huge increment of EER even for 
very low $\varepsilon$ values. FGSM and BIM attacks
with $\varepsilon=0.0001$ had a larger impact compared to the speaker identification task.
PWG defense increased benign EER by 40\% and reduced $\varepsilon=0.0001$ FGSM and BIM by 40 and 50\%, respectively.
For higher FGSM $\varepsilon$ and CW attacks, EER reduction was smaller (5-20\%). For high BIM with $\varepsilon\ge0.001$, the defense
did not have any effect. When we added randomized smoothing before the PWG, 
EER of attacked samples was reduced to the range 20-27\%, except for BIM $\varepsilon\ge 0.01$. This was at the cost of
increasing benign EER to 21.29\%. 

In summary, the results show that speaker verification is very vulnerable to attacks, as we showed in our
previous work~\cite{villalba2020x}.
The fully differentiable PWG defense provided significant protection for attacks with very low perceptibility.
However, we needed to add randomized smoothing to improve performance with higher $\varepsilon$, which significantly damaged
the benign performance.


\section{Conclusion}
\label{sec:conclusions}

We studied the robustness of several x-vector speaker identification systems to black-box and adaptive white-box adversarial attacks and analyzed four pre-processing defenses to increase adversarial robustness. Table~\ref{tab:results_compare} summarizes the results for the best configurations of each defense for all attacks. For easy visualization, we convert the table to a box plot (Figure~\ref{fig:results_boxplot}). The proposed defenses degraded benign accuracies very little--4.8 \% absolute in the worse case. This is a very mild degradation compared to the robustness when the system is adversarially attacked.
We observed that undefended systems were inherently robust to FGSM and Black-box Universal attacks transferred from a
speaker identification system with different architecture. However, they were extremely vulnerable under Iterative FGSM and CW attacks. 

The simple randomized smoothing defense robustified the system for imperceptible BIM ($L_\infty\le 0.01$) and CW recovering classification accuracies $\sim 97\%$. We also explored defenses based on three types of generative models: DefenseGAN, VAE, and PWG vocoder. These models learn the manifold of benign samples and were used to project adversarial examples (outside of the manifold) back into the clean manifold. DefenseGAN and VAE worked at the log-spectrogram level, while PWG worked at the waveform level. We noted that DefenseGAN performed poorly.
For VAE and PWG, we considered two configurations to create attacks that adapt to the defense: the adversary having access to an end-to-end differentiable (E2ED) version of the defense; or approximating the defense gradient using BPDA. In the BPDA setting, both VAE and PWG significantly improved adversarial robustness w.r.t. the undefended system. PWG performed the best, being close to the clean condition. For BIM attacks, PWG improved average accuracy from 17.16\% to a staggering 97.63\%. For CW, it improved from 1.4\% to 98.8\%. However, for the E2ED adaptive attack, VAE and PWG protection dropped significantly--PWG accuracy dropped to 37\% for CW attack.

Finally, we experimented with combining randomized smoothing with VAE or PWG. We found that smoothing followed by PWG vocoder was the most effective defense.  This defense protected against all BPDA attacks, including perceptible attacks with high $L_\infty$ perturbations--average accuracy was 93\% compared to 52\% in the undefended system. For adaptive attacks, accuracy only degraded for perceptible perturbations with $L_\infty > 0.01$.

We also tested our best defenses, namely PWG and smoothing before PWG on a speaker verification task. We used a fully differentiable version of PWG. We observed that PWG improved performance for attacks with very low perceptibility.
The fully differentiable PWG defense provides significant protection for attacks with very low perceptibility. However, we needed to add randomized smoothing to improve performance with higher $\varepsilon$, which damaged
the benign performance significantly.

In the future, we plan to evaluate these defenses in other speech tasks like automatic speech recognition tasks. Also, we plan to study how to improve the robustness of generative pre-processing defenses in the white-box setting. Finally, we are also interested in evaluating the merit of the proposed defenses for over-the-air attacks simulated using room impulse response as proposed by Qin et al.~\cite{Qin2019}.





\bibliographystyle{IEEEtran}

\bibliography{xvec_adv}

\setcounter{table}{0}
\renewcommand{\thetable}{A\arabic{table}}

\begin{table*}[t]
 \renewcommand{\arraystretch}{1}
    \centering
    \caption{\label{tab:vae_encdec} ResNet2d VAE encoder/decoder architectures. The first dimension of the input shows number of filter-banks, and the third dimension indicates the number of frames $T$. Batchnorm and ReLU activations were used after each convolution. Subpixel convolutions were used for upsampling operations.}
    \vspace{-2mm}
    \setlength\tabcolsep{3pt}
    \begin{tabular}{@{}lcccc@{}}
        \toprule
        & \multicolumn{2}{c}{\bf Encoder} &  \multicolumn{2}{c}{\bf Decoder} \\
        \cmidrule(rl){2-3}\cmidrule(l){4-5}
        \textbf{Layer name}   & \textbf{Structure}  & \textbf{Output} &  \textbf{Structure}  & \textbf{Output}  \\
        \midrule
        Input                 & --                          & $80 \times 1 \times T$  & 
        -- & $10 \times 80 \times T/8$ \\
        Conv2D-1              & $5 \times 5$, 64, Stride 1      & $80 \times 64 \times T$ 
        & $3 \times 3$, 512, Stride 1      & $10 \times 512 \times T/8$ \\
        \midrule
        ResNetBlock-1         & $\begin{bmatrix} 3 \times 3, 64  \\ 3 \times 3, 64  \end{bmatrix} \times 2$  , Stride 1& $80\times 64 \times T$ &
        $\begin{bmatrix} 3 \times 3, 512 \\ 3 \times 3, 512 \end{bmatrix} \times 2$, Stride $1$ & $10  \times 512  \times T/8$ \\
        ResNetBlock-2         & $\begin{bmatrix} 3 \times 3, 128  \\ 3 \times 3, 128  \end{bmatrix} \times 2$, Stride $2$ & $40 \times 128 \times T/2$ 
         & $\begin{bmatrix} 3 \times 3, 256 \\ 3 \times 3, 256 \end{bmatrix} \times 2$, Stride $2$ & $20 \times 256  \times T/4$\\
        ResNetBlock-3         & $\begin{bmatrix} 3 \times 3, 256 \\ 3 \times 3, 256 \end{bmatrix} \times 2$, Stride $2$ & $20 \times 256  \times T/4$ &
        $\begin{bmatrix} 3 \times 3, 128  \\ 3 \times 3, 128  \end{bmatrix} \times 2$, Stride $2$ & $40 \times 128 \times T/2$\\
        ResNetBlock-4         & $\begin{bmatrix} 3 \times 3, 512 \\ 3 \times 3, 512 \end{bmatrix} \times 2$, Stride $2$ & $10  \times 512  \times T/8$ &
        $\begin{bmatrix} 3 \times 3, 64  \\ 3 \times 3, 64  \end{bmatrix} \times 2$  , Stride 2& $80\times 64 \times T$ \\
        \midrule
        Projection & $1\times 1$, 80, Stride 1 & $10 \times 80 \times T/8$ & $1\times 1$,  1, Stride 1 & $80 \times 1 \times T$\\
        \bottomrule
    \end{tabular}
\end{table*}

\appendix[Neural network architectures]  

%



\begin{table}
 \renewcommand{\arraystretch}{1}
    \centering
    \caption{\label{tab:defgan_archi} GAN generator/discriminator architectures. For generator, latent\_dim =100 and ReLU activations were used after each deconvolution layer. For Discriminator,  LeakyReLU activations with negative slope = 0.2 were used after each convolution layer. For both, the output padding is $1 \times 1$}
    \vspace{-2mm}
    \setlength\tabcolsep{3pt}
    \begin{tabular}{@{}lcc@{}}
        \toprule
        & \multicolumn{2}{c}{\bf Generator} \\
        \cmidrule(lr){2-3}
        \textbf{Layer name}   & \textbf{Structure}  & \textbf{Output} \\
        \midrule
        Input & - &  latent\_dim\\
        Linear & - & $16384$\\
        Reshape & - & $1024 \times 4 \times 4$\\
        ConvTranspose2d-1 & $3 \times 3$ , 1024 , Stride 3 & $512 \times 7 \times 7$ \\
        ConvTranspose2d-2 & $3 \times 3$ , 512 , Stride 3 & $256 \times 13 \times 13$ \\
        ConvTranspose2d-3 & $3 \times 3$ , 256 , Stride 3 & $128 \times 25 \times 25$ \\
        ConvTranspose2d-4 & $3 \times 3$ , 128 , Stride 3 & $64 \times 49 \times 49$ \\
        ConvTranspose2d-5 & $3 \times 3$ , 64 , Stride 3 & $1 \times 97 \times 97$ \\
        Crop center & - & $1 \times 80 \times 80$ \\
        \midrule
        \midrule
         & \multicolumn{2}{c}{\bf Discriminator} \\
         \cmidrule(lr){2-3}
        \textbf{Layer name}   & \textbf{Structure}  & \textbf{Output} \\
        Input & - & $1 \times 80 \times 80$\\
        Conv2d-1 & $3 \times 3$ , 64 , Stride 2 & $64 \times 40 \times 40$ \\
        Conv2d-2 & $3 \times 3$ , 128 , Stride 2 & $128 \times 20 \times 20$ \\
        Conv2d-3 & $3 \times 3$ , 256 , Stride 2 & $256 \times 10 \times 10$ \\
        Conv2d-4 & $5 \times 5$ , 512 , Stride 2 & $512 \times 4 \times 4$ \\
        Conv2d-5 & $5 \times 5$ , 1024 , Stride 2 & $1024 \times 1 \times 1$ \\
        Reshape & - & 1024 \\
        Linear & - & 1 \\
        \bottomrule
    \end{tabular}
    \vspace{-4mm}
\end{table}

\begin{table}
  \renewcommand{\arraystretch}{1}
    \centering
    \caption{\label{tab:resnet34} ThinResNet34 x-vector architecture. $N$ in the last row is the number of speakers. The first dimension of the input shows number of filter-banks and the third dimension indicates the number of frames $T$.}
    \vspace{-2mm}
    \setlength\tabcolsep{2pt}
    \begin{tabular}{@{}lcc@{}}
        \toprule
        \textbf{Layer name}   & \textbf{Structure}          & \textbf{Output} \\
        \midrule
        Input                 & --                          & $80 \times 1 \times T$  \\
        Conv2D-1              & 3 $\times$ 3, 16, Stride 1      & $80 \times 16 \times T$ \\
        \midrule
        ResNetBlock-1         & $\begin{bmatrix} 3 \times 3, 16  \\ 3 \times 3, 16  \end{bmatrix} \times 3$  , Stride 1& $80\times 16 \times T$  \\
        ResNetBlock-2         & $\begin{bmatrix} 3 \times 3, 32  \\ 3 \times 3, 32  \end{bmatrix} \times 4$, Stride $2$ & $40 \times 32 \times T/2$  \\
        ResNetBlock-3         & $\begin{bmatrix} 3 \times 3, 64 \\ 3 \times 3, 64 \end{bmatrix} \times 6$, Stride $2$ & $20 \times 64  \times T/4$ \\
        ResNetBlock-4         & $\begin{bmatrix} 3 \times 3, 128 \\ 3 \times 3, 128 \end{bmatrix} \times 3$, Stride $2$ & $10  \times 128  \times T/8$ \\
        \midrule
        Flatten               & --                & $1280 \times T/8$                            \\
        StatsPooling          & --                & $2560$                 \\
        \midrule
        Dense1                & --                & $256$                            \\
        Dense2 (Softmax)      & --                & $N$                              \\
        \bottomrule
    \end{tabular}
    \vspace{-4mm}
\end{table}


Table~\ref{tab:vae_encdec} describes the Encoder/Decoder architectures for the VAE defense.

Table~\ref{tab:defgan_archi}, shows the architectures for the DefenseGAN generator and discriminator. 
The generator projects the latent variable $\zvec$ with dimension 100 to $256\times d$ dimension and reshape to create a tensor of size
$16\times 16$ with $16\times d$ channels. A sequence of 2$D$ transposed convolutions with stride 2 upsamples this tensor to generate square log-Mel-filter-bank patches with 80 frames of dimension 80. The critic is a sequence of 2$D$ convolutions with stride 2, which downsamples the input to generate a single output per filter-bank patch. We did not observe improvements by increasing the complexity of generator and discriminator networks. The DefenseGAN procedure used a 0.025 learning rate, 300 iterations, and 10 random seeds at inference time. 

Table~\ref{tab:resnet34} describes the architecture of the ThinResNet34 x-vector model, which we use to compare multiple defenses.

\end{document}